\pdfoutput=1
\documentclass[a4paper,11pt]{article}
\usepackage{jcappub}
\usepackage[utf8]{inputenc}

\usepackage{amsmath,amssymb}
\usepackage{graphicx}
\usepackage{multicol}
\usepackage{hyperref}
\usepackage{physics}
\usepackage{lipsum}
\usepackage{cancel}

\def\a{\alpha}
\def\b{\beta}

\def\d{\delta}

\def\g{\gamma}

\def\k{\kappa}
\def\l{\lambda}
\def\m{\mu}
\def\n{\nu}

\def\r{\rho}
\def\s{\sigma}

\def\u{\upsilon}
\def\x{\xi}

\def\ta{\tilde{\a}}
\def\tb{\tilde{\b}}

\usepackage{scalerel}
\usepackage{tikz}
\usetikzlibrary{svg.path}

\definecolor{orcidlogocol}{HTML}{A6CE39}
\tikzset{
  orcidlogo/.pic={
    \fill[orcidlogocol] svg{M256,128c0,70.7-57.3,128-128,128C57.3,256,0,198.7,0,128C0,57.3,57.3,0,128,0C198.7,0,256,57.3,256,128z};
    \fill[white] svg{M86.3,186.2H70.9V79.1h15.4v48.4V186.2z}
                 svg{M108.9,79.1h41.6c39.6,0,57,28.3,57,53.6c0,27.5-21.5,53.6-56.8,53.6h-41.8V79.1z M124.3,172.4h24.5c34.9,0,42.9-26.5,42.9-39.7c0-21.5-13.7-39.7-43.7-39.7h-23.7V172.4z}
                 svg{M88.7,56.8c0,5.5-4.5,10.1-10.1,10.1c-5.6,0-10.1-4.6-10.1-10.1c0-5.6,4.5-10.1,10.1-10.1C84.2,46.7,88.7,51.3,88.7,56.8z};
  }
}

\newcommand\orcid[1]{\href{https://orcid.org/#1}{\mbox{\scalerel*{
\begin{tikzpicture}[yscale=-1,transform shape]
\pic{orcidlogo};
\end{tikzpicture}
}{|}}}}

\newcommand{\spc}{\,\,\,\,} 
\newcommand{\be}{\begin{equation}} 
\newcommand{\ee}{\end{equation}}
\newcommand{\beq}{\begin{equation}} 
\newcommand{\eeq}{\end{equation}}
\newcommand{\bea}{\begin{equation}\begin{aligned}} 
\newcommand{\eea}{\end{aligned}\end{equation}}
\newcommand{\ba}{\begin{eqnarray}}
\newcommand{\ea}{\end{eqnarray}}

\usepackage{csquotes}
\definecolor{tclr}{RGB}{103,103,246}
\def\eps{\epsilon}

\def\ubar{\bar{U}}

\title{Scale-Invariant Quadratic Gravity and Inflation in the Palatini Formalism}

\author[a]{Ioannis D.~Gialamas~\orcid{0000-0002-2957-5276},}
\author[b]{Alexandros Karam~\orcid{0000-0002-0582-8996},}
\author[c]{Thomas D.~Pappas~\orcid{0000-0003-2186-357X}}
\author[a]{and Vassilis C.~Spanos~\orcid{0000-0001-8676-3655}}

\emailAdd{i.gialamas@phys.uoa.gr}
\emailAdd{alexandros.karam@kbfi.ee} 
\emailAdd{thomas.pappas@physics.slu.cz}
\emailAdd{vspanos@phys.uoa.gr}

\vspace{5cm}

\affiliation[a]{ National and Kapodistrian University of Athens, Department of Physics, \\
 Section of Nuclear {\rm \&} Particle Physics,  GR--157 84 Athens, Greece }
\affiliation[b]{Laboratory of High Energy and Computational Physics, 
National Institute of Chemical Physics and Biophysics, R{\"a}vala pst.~10, Tallinn, 10143, Estonia}
\affiliation[c]{Research Centre for Theoretical Physics and Astrophysics, Institute of Physics, Silesian University in Opava, Bezručovo nám.~13, CZ-746 01 Opava, Czech Republic}

\abstract{In the framework of classical scale invariance, we consider quadratic gravity in the Palatini formalism and investigate the inflationary predictions of the theory. Our model corresponds to a two-field scalar-tensor theory, that involves the Higgs field and an extra scalar field stemming from a gauge $U(1)_X$ extension of the Standard Model, which contains an extra gauge boson and three right-handed neutrinos. Both scalar fields couple nonminimally to gravity and induce the Planck scale dynamically, once they develop vacuum expectation values. By means of the Gildener-Weinberg approach, we describe the inflationary dynamics in terms of a single scalar degree of freedom along the flat direction of the tree-level potential. The one-loop effective potential in the Einstein frame exhibits plateaus on both sides of the minimum and thus the model can accommodate both small and large field inflation. The inflationary predictions of the model are found to comply with the latest bounds set by the Planck collaboration for a wide range of parameters and the effect of the quadratic in curvature terms is to reduce the value of the tensor-to-scalar ratio.}

\begin{document}

\maketitle

\vspace{1 cm}

\section{Introduction} \label{sec:intro}

The combined analysis of the latest cosmological data based on various observations such as the cosmic microwave background (CMB), the large scale structures, the supernova data etc., favor~\cite{Aghanim:2018eyx} a flat, homogeneous and isotropic Universe. Cosmic inflation~\cite{Starobinsky1980, Guth1981, Linde1982, Albrecht1982} not only naturally explains the aforementioned features of the Universe, but quite importantly, when treated quantum mechanically, it also provides a mechanism for the production of the necessary primordial anisotropies, which act as seeds for the generation of the large scale structures that we observe today. Data from the Planck mission combined with previous observations~\cite{Akrami:2018odb} have severely constrained the parameter space of the inflationary models and essentially ruled out many of those, including the simplest ones where a scalar field is minimally coupled to gravity. On the other hand, more involved models such as the Starobinsky~\cite{Starobinsky1980}, where an $\mathcal{R}^2$ term is added to the Einstein-Hilbert action, seem to lie within the allowed range. This kind of nonminimal models belongs in the general class of scalar-tensor (ST) theories~\cite{Faraoni:1998qx, Faraoni2004a, Flanagan:2004bz, Jarv:2007iq, Chiba:2013mha, Postma:2014vaa, Jarv:2014hma, Jarv:2015kga, Kuusk:2016rso, Jarv:2016sow, Karam:2017zno, Burns2016, Karamitsos:2018lur, Jarv:2020qqm, Gialamas:2020vto, Karam:2021wzz}. In such models, the scalar field $\phi$ typically couples to gravity via a term of the form $\xi \phi^2 \mathcal{R}$,  where $\xi$ is a dimensionless coupling constant and $\mathcal{R}$ is the Ricci scalar. It is noteworthy that this type of coupling allows for the Planck scale to be generated dynamically when $\phi$ develops a vacuum expectation value (VEV).
 
The dynamical generation of the Planck scale is usually achieved in scale-invariant theories~\cite{Shaposhnikov2009a, GarciaBellido:2011de, Bezrukov:2012hx, Khoze2013, Steele:2013fka, Ren:2014sya,  Kannike2014, Csaki2014, Kannike2015a,  Kannike2016b, Wang:2015cda, Barvinsky:2015uxa, Farzinnia:2015fka, Rinaldi2016a, Marzola2016b, Barrie2016, Ferreira2016, Kannike2017a, Marzola2016, Karananas2016, Tambalo2017, Kannike2017, Artymowski2017, Ferreira:2016wem, Salvio2017, Karam2017, Kaneta2017, Karam2017a, Racioppi2018, Ferreira:2018a, Benisty:2018fja, Barnaveli:2018dxo, Kubo:2018kho, Mooij:2018hew, Shaposhnikov:2018nnm, Wetterich:2019qzx, Vicentini:2019etr, Shkerin:2019mmu, Ferreira:2019zzx, Ghilencea:2019rqj, Salvio:2019wcp, Oda:2019iwc, Racioppi:2019jsp, Benisty:2020nuu, Benisty:2020vvm, Tang:2020ovf, Gialamas:2020snr, Ghilencea:2020rxc}, where the running of the inflaton quartic coupling induces symmetry breaking \`{a} la Coleman--Weinberg. Scale invariance posits that the Lagrangian of a theory should not contain any ad hoc mass parameters. Utilizing the restrictive power of scale invariance, one can built three more terms that respect the symmetry: the Starobinsky term $\alpha \mathcal{R}^2$ and the terms $\beta \mathcal{R}_{\mu\nu} \mathcal{R}^{\mu\nu}$ and $\gamma \mathcal{R}_{\mu\nu\sigma\lambda} \mathcal{R}^{\mu\nu\sigma\lambda}$, where $\mathcal{R}_{\mu\nu\sigma\lambda}$ and $\mathcal{R}_{\mu\nu}$ are the Riemann and Ricci tensors, respectively, and $\alpha$, $\beta$ and $\gamma$ are dimensionless constants. This gravitational theory is called {\textit{quadratic gravity}} and has recently received a lot of attention as a possible realization of quantum gravity~\cite{Stelle:1976gc, Biswas:2005qr, Salvio2014, Edery:2014nha,  Farzinnia:2015fka, Salvio2018, Salvio2017, Salvio:2018crh, Salvio:2018kwh,  Salvio:2019wcp, Edery:2019bsh, Ghilencea:2019rqj, Salvio:2019llz, Ghilencea:2020piz,  Ghilencea:2020rxc,  Salvio:2020axm}. Of course, in extended theories of gravity, the issue of the correct formulation arises, i.e. whether one should employ the metric or the Palatini formalism when varying the action.
    
It has  been known that the Palatini formulation~\cite{Palatini1919, Ferraris1982} of General Relativity (GR) (first-order formalism) is an alternative to the well-known metric formulation (second-order formalism). In the latter, the spacetime connection is the usual Levi-Civita one, while in the Palatini approach the connection $\Gamma^\l_{\,\,\m\n}$ and the metric $g_{\m\n}$  are treated as independent variables. In the context of GR, the two formalisms are equivalent at the level of the field equations, with the Levi-Civita connection in the Palatini approach being recovered on shell. When nonminimal couplings between gravity and matter~\cite{Bauer:2008zj, Bauer:2010bu, Tamanini:2010uq, Bauer:2010jg, Rasanen:2017ivk, Tenkanen:2017jih, Racioppi:2017spw, Markkanen:2017tun, Jarv:2017azx, Fu:2017iqg, Racioppi:2018zoy, Carrilho:2018ffi, Kozak:2018vlp, Rasanen:2018fom, Rasanen:2018ihz, Almeida:2018oid, Shimada:2018lnm, Takahashi:2018brt, Jinno:2018jei, Rubio:2019ypq, Bostan:2019uvv, Bostan:2019wsd, Tenkanen:2019xzn, Racioppi:2019jsp, Tenkanen:2020dge, Shaposhnikov:2020fdv, Borowiec:2020lfx, Jarv:2020qqm, Karam:2020rpa, McDonald:2020lpz, Langvik:2020nrs, Shaposhnikov:2020gts, Shaposhnikov:2020frq, Gialamas:2020vto, Mikura:2020qhc, Verner:2020gfa, Enckell:2020lvn, Reyimuaji:2020goi, Karam:2021wzz, Mikura:2021ldx, Kubota:2020ehu, Saez-ChillonGomez:2021byq} or/and $f(R)$ theories\footnote{Throughout this paper we use different symbols for the curvature scalar and tensors, which in the metric formulation we denote by $\mathcal{R}$, while in the Palatini approach by $R$.}~\cite{Olmo:2011uz, Bombacigno:2018tyw, Enckell:2018hmo, Antoniadis:2018ywb, Antoniadis:2018yfq, Tenkanen:2019jiq, Edery:2019txq, Giovannini:2019mgk, Tenkanen:2019wsd, Gialamas:2019nly, Tenkanen:2020cvw, Lloyd-Stubbs:2020pvx, Antoniadis:2020dfq, Ghilencea:2020piz, Das:2020kff, Gialamas:2020snr, Ghilencea:2020rxc, Bekov:2020dww, Dimopoulos:2020pas, Gomez:2020rnq, Karam:2021sno, Lykkas:2021vax} are considered, the resultant field equations are no longer the same and thus the two formalisms lead to different cosmological predictions. A remarkable example is the Starobinsky model of inflation~\cite{Starobinsky1980}, where the addition of an $\mathcal{R}^2$ term in the usual Einstein-Hilbert action is translated to a new propagating scalar degree of freedom which plays the role of the inflaton. In the Palatini formalism there are no extra propagating degrees of freedom, therefore the inflaton has to be added ad hoc in the action. The advantage of considering the Palatini formulation is that the addition of the $R^2$ term can be used to reduce the tensor-to-scalar ratio $r$~\cite{Enckell:2018hmo}. Thereby, various models where inflation is driven by a scalar field can be rendered again compatible with the observations~\cite{Antoniadis:2018ywb, Antoniadis:2018yfq}.
Furthermore, the addition of a symmetric Ricci tensor squared term $R_{(\m\n)}R^{(\m\n)}$ in the Einstein-Hilbert action has the same effect as the pure $R^2$ term (see~\cite{Annala2020}), at least with respect to the modification of the scalar potential, and consequently leads to the reduction of the tensor-to-scalar ratio~\cite{Enckell:2018hmo}. The main, but not significant, difference between these two quadratic scale-invariant terms is that in the Einstein frame (EF) the $R^2$ term translates also to a second-order kinetic term, while the $R_{(\m\n)}R^{(\m\n)}$ term yields a series of higher-order kinetic terms. These higher-order kinetic terms are nevertheless negligible at least during slow roll. 

In this paper we construct a model of scale-invariant quadratic gravity, where the Planck scale is dynamically generated through the VEVs of a scalar field $\phi$ and the Higgs $h$, which are nonminimally coupled to gravity through terms of the form $\xi_i \Phi_i^2 R$, where $\Phi_i = \phi, \, h$. The extra scalar field $\phi$ stems from a general $U(1)_X$ extension of the Standard Model (SM) gauge structure that contains an extra gauge boson $X_\mu$ and three right-handed neutrinos $N_R$ that can generate masses for the SM neutrinos via a type-I seesaw mechanism. The model can easily accommodate dark matter in a natural way and we outline three distinct possibilities. Moreover, the mass of the Higgs and the electroweak scale is generated through a portal coupling between $\phi$ and $h$ of the form $\lambda_{h \phi} h^2 \phi^2$. Thus, the addition of the extra scalar field $\phi$ is necessary to preserve the scale invariance of our model since the known Higgs mass term contained in the SM Lagrangian is not scale invariant.

The rest of the paper is organised as follows: In Section~\ref{sec:QuadGrav}, we describe the beyond SM (BSM) part of the model, along with the extended gravity part that contains terms quadratic in curvature and is studied in the Palatini formulation. We also briefly describe the dark matter candidates that can arise from our setup. Then, in Section~\ref{sec:GW_formalism}, we employ the Gildener-Weinberg approach~\cite{Gildener1976a} (which is a generalization of the Coleman-Weinberg mechanism~\cite{Coleman1973} to multiple fields) in order to obtain the flat direction of the tree-level potential. Along the flat direction, the theory effectively becomes single field and by computing the quantum corrections we obtain the one-loop effective potential, which is stable due to the extra $U(1)_X$ gauge boson. In Section~\ref{sec:EF_rep}, we introduce an auxiliary field $\Sigma_{\mu\nu}$ in order to parametrize the terms quadratic in curvature. By applying a Weyl rescaling and a disformal transformation of the metric, we obtain the EF representation of the theory. In the process, higher-order kinetic terms arise, but these have been shown to not significantly influence the inflationary~\cite{Tenkanen:2020cvw} and reheating~\cite{Karam:2021sno} dynamics. At the same time, the EF potential is modified and develops plateaus on both sides of the VEV. In Section~\ref{sec:Inflation}, we obtain the inflationary predictions of the model in the slow roll approximation and impose constraints on the free parameters. Finally, we summarize and conclude in Section~\ref{sec:Conclusions}. Further details about the disformal transformation are relegated to the Appendices.

We use natural units $\hbar = c = k_\mathrm{B} = 1$ and the metric signature $(-,\!+,\!+,\!+)$ throughout. We also use $M_{\rm P}^2 = 1$ in most formulas except when we want the dimensionality to be explicit.

%%%%%%%%%%%%%%%%%%%%%%%%%%%%%%%%%%%%%%%%%%%%%%
\section{The model}
\label{sec:QuadGrav}

We begin the discussion of our model by describing the scale-invariant $U(1)_X$
extension of the SM that we use~\cite{Hempfling1996, Chang:2007ki, Iso2009, Iso2009a, Iso2013, Englert2013, Chun:2013soa, Khoze2013a, Radovcic:2014rea, Khoze2014, Benic:2014aga, Guo:2015lxa, Wang:2015sxe, Altmannshofer2015, Das:2015nwk, Jinno:2016knw, Das2016, Oda:2017kwl, Marzo:2018nov}, which contains a complex scalar field $\Phi$, a gauge boson $X_\mu$ and three right-handed neutrinos $N^i_R$. We also outline three distinct possibilities for 
the model to accommodate dark matter candidates. Subsequently, we focus on the gravity part of the theory and study it in the Palatini formalism.

\subsection{$U(1)_X$ extension of the Standard Model}

We consider the $U(1)_X$ extension of the SM based on the gauge group $SU(3)_c \cross SU(2)_L \cross U(1)_Y \cross U(1)_X$. In Table~\ref{Table:1} we present the matter fields of this model which contains the SM matter fields along with three generations of right-handed neutrinos $N^{i}_{R}$ ($i=1,2,3$) and a $U(1)_X$ complex scalar field $\Phi$, whose VEV will generate the mass of the vector boson $X_\m$ as well as the masses of the right-handed neutrinos. This $U(1)_X$ extension can be recognized as a linear combination of the $U(1)_Y$  and the $U(1)_{B-L}$ gauge group, with the latter being free of gauge and gravitational anomalies. The existence of the three right-handed neutrinos plays a crucial role to this anomaly cancellation. Following~\cite{Oda2018} we introduce the real parameters $x_H$ and $x_\Phi$ which are used in the determination of the $U(1)_X$ charge of the field $\Phi$, that is given by
\be
Q_X =  Y x_H + Q_{BL}  \; x_\Phi\,,
\label{eq:Qcharge}
\ee
with $Y$ and $Q_{BL}$ being its hypercharge and $B-L$ charge respectively. Two interesting choices for the parameters $x_H$ and $x_\Phi$ are the choice  $(x_H,x_\Phi)=(0,1)$, which corresponds to the $U(1)_{B-L}$ model and the $(x_H, x_\Phi) = (-2,1)$ choice which coincides with the SM with an additional $U(1)_{R}$ symmetry. 

The covariant derivative associated with the $U(1)_Y \cross U(1)_X$ gauge interaction is defined as
\be
D_\m  =  \partial_\m -i\left(g_1 Y+\tilde{g}Q_X\right)B_\m -i g_X Q_X X_\m \,,
\label{eq:covder}     
\ee
where $g_1$ and $g_X$ are the $U(1)_Y$ and $U(1)_X$ gauge couplings respectively. In~\eqref{eq:covder}, the possible kinetic mixing between the two $U(1)$ gauge bosons can be ignored for simplicity assuming that the mixing coupling $\tilde{g}$ vanishes at the $U(1)_X$ symmetry breaking scale.
%%%%%%%%%%%%%%%%%%%%%%%%%%%%%%%%%%%%%%
\begin{table}[h]
\begin{center}
\begin{tabular}{|c|ccc|c|}
\hline
      &  $SU(3)_c$  & $SU(2)_L$ & $U(1)_Y$ & $U(1)_X$  \\ 
\hline
$q^{i}_{L}$ & {\bf 3 }    &  {\bf 2}         & $ 1/6$       & $(1/6) x_{H} + (1/3) x_{\Phi}$   \\
$u^{i}_{R}$ & {\bf 3 }    &  {\bf 1}         & $ 2/3$       & $(2/3) x_{H} + (1/3) x_{\Phi}$   \\
$d^{i}_{R}$ & {\bf 3 }    &  {\bf 1}         & $-1/3$       & $(-1/3) x_{H} + (1/3) x_{\Phi}$  \\
\hline
$\ell^{i}_{L}$ & {\bf 1 }    &  {\bf 2}         & $-1/2$       & $(-1/2) x_{H} +(-1) x_{\Phi}$    \\
$e^{i}_{R}$    & {\bf 1 }    &  {\bf 1}         & $-1$                   & $(-1) x_{H} +(-1) x_{\Phi}$   \\
\hline
$H$            & {\bf 1 }    &  {\bf 2}         & $ 1/2$       & $(1/2) x_{H}$   \\  
\hline
$N^{i}_{R}$    & {\bf 1 }    &  {\bf 1}         &$0$                    & $(-1) x_{\Phi}$     \\
$\Phi$            & {\bf 1 }       &  {\bf 1}       &$ 0$                  & $ (+ 2) x_{\Phi}$  \\ 
\hline
\end{tabular}
\end{center}
\caption{
\sf The matter fields of the $U(1)_X$ extension of the SM along with the corresponding charges. In addition to the SM particle content ($i=1,2,3$), three right-handed neutrinos $N_R^i$ ($i=1, 2, 3$) and a $U(1)_X$ complex scalar field $\Phi$ are introduced. The $U(1)_X$ charge is determined by the two real parameters, $x_H$ and $x_\Phi$, as  $Q_X =  Y x_H + Q_{BL}  \, x_\Phi$ with its hypercharge $Y$ and $B-L$ charge $Q_{BL}$.
}
\label{Table:1}
\label{table}
\end{table}
%%%%%%%%%%%%%%%%%%%%%%%%%%%%%%%%%%%%%%%%%%%%%%%

In the known SM Yukawa sector we need to add the BSM Yukawa sector arising from the $U(1)_X$ extension which reads
\be
\label{eq:Yukawa}
\mathcal{L}^{\rm BSM}_{\rm Yukawa} = -y_D^{i j} \overline{\ell}^{i}_{L} H N^{j}_{R} -\frac{1}{2} y_M^{i} \Phi  \overline{N}^{i C}_{R} N^{i}_{R} + h.c.\,,
\ee
where $y_D$ and $y_M$  are the Dirac and Majorana Yukawa couplings respectively. Also, without loss of generality, the Majorana Yukawa couplings are assumed to be already diagonal in our basis. Furthermore, it is interesting to note that in this setting a lepton asymmetry can be generated from decays of the heavy right-handed neutrinos into SM leptons at high temperatures. Then, the lepton asymmetry can be converted into a baryon asymmetry via electroweak sphalerons~\cite{Davidson:2002qv, Davidson:2008bu} (see also~\cite{Khoze2013a, Khoze2016, Biswas:2017tce}).

Assuming that the $U(1)_X$ complex scalar field $\Phi$ develops a nonzero VEV $v_\phi$ and working in the unitary gauge, we have that
\be
\label{eq:unit_gauge}
\Phi = \frac{1}{\sqrt{2}}(\phi + v_\phi)\,.
\ee
Thus, the BSM scalar Lagrangian and the gravity Lagrangian are given by
\ba
\label{eq:scal_grav}
\mathcal{L}^{\rm BSM}_{\rm scalar} &=&  -\frac{1}{2} g^{\m\n} \partial_\m \phi \partial_\n \phi - \frac{1}{4} \l_\phi \phi^4 +\frac{1}{4} \l_{h \phi} h^2\phi^2\,, \nonumber
\\ \mathcal{L}_{\rm gravity} &=&  \frac{1}{2} \left( \xi_\phi \phi^2 + \xi_h h^2 \right) g^{\mu\nu} R_{\mu\nu} (\Gamma) \,,
\ea
where $h$ is the Higgs field also written in the unitary gauge and $\xi_\phi$, $\xi_h$ are the nonminimal couplings between gravity and matter. Note that the Ricci tensor depends only on the connection $\Gamma$ since we are working in the Palatini formalism. Also, there are no mass terms for either $\phi$ or $h$ since the theory must respect classical scale invariance. The reduced Planck mass $M_{\rm P}$ is generated dynamically when $\phi$ and h develop their VEVs,
\be
\label{eq:planckscale}
M^2_{\rm P} = \xi_\phi v_\phi^2 + \xi_h v_h^2\,.
\ee
Associated with the $U(1)_X$ and the electroweak symmetry breaking, the $U(1)_X$ gauge boson $X_\mu$ and the Majorana right-handed neutrinos $N^i_R$ acquire their masses as
\be 
M_X = \sqrt{ \left( 2 x_\Phi g_X v_\phi \right)^2 + \left( x_H g_X v_h \right)^2} \simeq 2 x_\Phi g_X v_\phi \,, \qquad M_{N^i_R} = \frac{y^i_M}{\sqrt{2}} v_\phi \,.
\ee

The part of the action that contains the scalar $\phi$ and the Higgs $h$ is
\be 
S = \int \dd^4 x \sqrt{-g} \left[ \frac{1}{2} \left( \xi_\phi \phi^2 + \xi_h h^2 \right) g^{\mu\nu} R_{\mu\nu} (\Gamma) -\frac{1}{2} g^{\mu\nu} \partial_\mu \phi \partial_\nu \phi - \frac{1}{2} g^{\mu\nu} \partial_\mu h \partial_\nu h - V^{(0)}(\phi,h) \right]\,,
\ee
with the tree-level potential given by
\be
\label{eq:scalar_pot}
 V^{(0)}(\phi,h) =\frac{1}{4}\left( \lambda_\phi \phi^4- \lambda_{h\phi} h^2 \phi^2 + \lambda_{h} h^4 \right) \,.
\ee
Note that the coupling constants $\lambda_\phi$, $\lambda_h$ and $\lambda_{h\phi}$ are dimensionless, assumed to be positive and the minus sign in front of the portal coupling term is introduced to allow for the spontaneous breaking of the symmetry due to the running of the coupling constants.

\subsection{Potential dark matter candidates}

An interesting feature of the $U(1)_X$ model under consideration is that it can provide us with viable dark matter candidates in a minimal and natural way. 
\begin{enumerate}
    \item A first possibility is that the extra gauge boson $X_\mu$ constitutes dark matter~\cite{Lebedev2012a, Baek2013, Gross2015}. The $U(1)_X$ gauge group contains an intrinsic $Z_2$ discrete symmetry,  which automatically renders $X_\mu$ stable. Note, however, that this statement applies if that $U(1)_X$ is sequestered and has no tree-level mixing with the hypercharge. In that case, no mixing can be
    generated at the one-loop level either. 
    
    \item A second possibility arises by introducing a $Z_2$ parity and imposing one of the three right-handed neutrinos to be odd, while the others are even~\cite{Okada:2010wd,Benic2015}. Thus, the odd right-handed neutrino becomes stable and can be a DM candidate. The rest of the right-handed neutrinos suffice to produce the observed neutrino oscillations.
    
    \item A third possibility arises by adding an extra Dirac fermion $\zeta$, which is singlet under the SM gauge group and has a generic $\mathrm{U}(1)_X$ charge $Q_X$~\cite{Kawai2021}. It is worth noting that the addition of the Dirac fermion does not spoil the anomaly cancellation of the $U(1)_X$ extended SM. The $\zeta$ field interacts with the SM particles due to $U(1)_X$ gauge interactions and its relic freeze-out abundance is calculated through the processes $\zeta \bar{\zeta}  \overset{X_\m}{\longleftrightarrow}  f \bar{f}$, where $f$ is a SM fermion. On the other hand in~\cite{Okada2020}, a freeze-in DM scenario is studied, where either $X_\m$ or the right-handed neutrinos can be light of the order $100\ \rm MeV$ to $1\ \rm GeV$. 
\end{enumerate}

\subsection{Palatini quadratic gravity}

In the metric formulation of gravity, the large number of symmetries of the Riemann tensor allows one to consider only a few quadratic terms, namely $\mathcal{R}^2$, $\mathcal{R}_{\mu\nu} \mathcal{R}^{\mu\nu}$ and $\mathcal{R}_{\mu\nu\sigma\lambda} \mathcal{R}^{\mu\nu\sigma\lambda}$. Furthermore, by virtue of the Gauss-Bonnet theorem the homonymous term $\mathcal{R}^2 -4\mathcal{R}_{\mu\nu} \mathcal{R}^{\mu\nu}+\mathcal{R}_{\mu\nu\sigma\lambda} \mathcal{R}^{\mu\nu\sigma\lambda}$ reduces to a topological surface term in four dimensions and thus solving for the quadratic in the Riemann tensor term one ends up with
\be 
S = \int \dd^4 x \sqrt{-g} \left( \frac{M^2_{\rm P}}{2} \mathcal{R} + \alpha \mathcal{R}^2 + \beta \mathcal{R}_{\mu\nu} \mathcal{R}^{\mu\nu}  \right) \,.
\ee
This Lagrangian has been proven to be renormalizable in all orders of perturbation theory~\cite{Stelle:1976gc}, but this comes with the cost of a ghostlike antigraviton state~\cite{Salvio2014, Salvio:2018crh}.

On the other hand, in the Palatini formulation the situation is slightly more complicated when adding quadratic curvature invariants to the action. In contrast to the metric case there is now a plethora of invariants that can be constructed out of the Ricci and Riemann tensors~\cite{Borunda:2008kf,Annala2020}. Actually there are three different nonvanishing contractions of the Riemann tensor, so the Ricci tensor can be defined as\footnote{Although the Ricci tensor is not unique, the Ricci scalar is and can be defined as $R=g^{\mu\nu}R_{\mu\nu}=g^{\mu\nu}\hat{R}_{\mu\nu}$, while the third possible contraction $g^{\mu\nu}R'_{\mu\nu}$ vanishes due to the symmetry of the metric tensor.}
\be
R_{\m\n}= R^\lambda_{\spc\m\lambda\n}, \spc \hat{R}^{\m}_{\spc\n}=g^{\lambda\sigma} R^\mu_{\spc\sigma\nu\lambda} \spc \text{and} \spc R'_{\m\n} = R^\lambda_{\spc\lambda\m\n}\,.
\label{eq:pos_Ricci}
\ee
The most general Lagrangian second order in the Riemann tensor contains 16 possible contractions and can be written as
\ba 
S &=& \int \dd^4 x \sqrt{-g} \left[  \alpha R^2 + \beta_1 R_{\m\n} R^{\m\n} + \beta_2 R_{\m\n} R^{\n\m} + \beta_3 R_{\m\n} \hat{R}^{\m\n} + \beta_4 R_{\m\n} \hat{R}^{\n\m}  + \right. \nonumber \\
  && \left. \qquad  \beta_5 \hat{R}_{\m\n} \hat{R}^{\m\n} + \beta_6 \hat{R}_{\m\n} \hat{R}^{\n\m}  + \beta_7 \hat{R}_{\m\n} R'^{\m\n}  + \beta_8 R'_{\m\n} R'^{\m\n}  + \beta_9 R_{\m\n} R'^{\m\n}  + \right. \nonumber \\
  && \left. \qquad  \gamma_1 R_{\m\n\sigma\lambda} R^{\m\n\sigma\lambda} + \gamma_2 R_{\m\n\sigma\lambda} R^{\m\sigma\n\lambda} + \gamma_3 R_{\m\n\sigma\lambda} R^{\n\m\sigma\lambda} + \gamma_4 R_{\m\n\sigma\lambda} R^{\n\sigma\m\lambda}  +  \right. \nonumber \\
  && \left. \qquad \gamma_5 R_{\m\n\sigma\lambda} R^{\sigma\n\m\lambda} + \gamma_6 R_{\m\n\sigma\lambda} R^{\sigma\lambda\m\n}
  \right]\,.
\ea
The quadratic action above is too complicated, therefore in order to make our computations analytically tractable, we will work under several simplifying assumptions. First, we assume that the spacetime connection is symmetric $\Gamma^\l_{\,\,\m\n} = \Gamma^\l_{\,\,\n\m}$, like the Levi-Civita one. Additionally, we discard the terms constructed from the Riemann tensor. In order to consider an invariant action under projective transformations, which  does  not  introduce  extra  gravitational degrees of freedom, we assume only a symmetric Ricci tensor. Nonsymmetric Ricci and therefore metric tensors contain new gravitational degrees of freedom and can lead to instabilities~\cite{Damour1993a}. Then, the action that we consider in the Jordan frame (JF) containing the %inflaton 
$\phi$, the Higgs $h$ and the nonminimal couplings between gravity and matter reads\footnote{From now on we consider only the symmetric Ricci tensor $R_{(\m\n)}$ and in order to speed up notation we discard the parentheses.}
\ba 
S &=&  \int \dd^4 x \sqrt{-g} \left\lbrace \frac{1}{2} \left[ \left( \xi_\phi \phi^2 + \xi_h h^2 \right) g^{\mu\nu} R_{\mu\nu} + \alpha R^2 + \beta R_{\mu\nu} R^{\mu\nu} \right] \right. \nonumber \\
&& \qquad \qquad \left. -\frac{1}{2} g^{\mu\nu} \partial_\mu \phi \partial_\nu \phi - \frac{1}{2} g^{\mu\nu} \partial_\mu h \partial_\nu h - V^{(0)}(\phi,h)\right\rbrace\,,
\label{eq:action_jf}
\ea
where the most general classically scale-invariant potential that can be constructed out of two real scalar fields is given by~\eqref{eq:scalar_pot}.
Notice that one could consider a doublet of inflatons without including the $R_{\mu\nu} R^{\mu\nu}$ term or add $R_{\mu\nu} R^{\mu\nu}$ without adding a doublet of inflatons. Nevertheless, we have included both extensions in the action for reasons of generality as well as for practical reasons associated with the lowering of the tensor-to-scalar ratio to observationally viable values and the dynamical generation of the Planck mass. More precisely, the specific choice of higher-curvature extension of the action is justified on the grounds of considering a general gravity sector that respects classical scale invariance and goes beyond the $R^2$ term that is known to lower the predicted value for the tensor-to-scalar ratio. This way, we will be in a position to investigate the interplay between the higher-curvature corrections and the overall effect that they have on the inflationary predictions. The inclusion of the extra scalar is deemed necessary for the dynamical generation of the Planck scale mainly via the VEV of  the extra scalar field in order to avoid an unnaturally large value for the nonminimal coupling constant of the Higgs field with gravity that would be otherwise required.

With the aim of eventually recasting the action~\eqref{eq:action_jf} in the EF where the gravity sector consists solely of the Einstein-Hilbert term, we will start by performing a  Weyl rescaling of the metric of the form 
\be
g^{\m\n} \longrightarrow \Omega ^2 g^{\m\n}\,,\spc \Omega ^2 =\x _\phi \phi ^2 + \x _h h ^2\,.
\label{eq:Weylresc}
\ee
The quadratic in curvature terms are invariant under the rescaling~\eqref{eq:Weylresc} in contrast to the Einstein-Hilbert term which rescales as $R \longrightarrow \Omega ^2 R$ and so the action takes the form
\ba
S =  \int \dd^4 x \sqrt{-g} && \left\lbrace \frac{1}{2} \left[  g^{\m\n} R_{\m\n} + \alpha R^2 + \beta R_{\m\n} R^{\m\n} \right] \right.  -\frac{1}{2\Omega ^2} g^{\m\n} \partial_\m \phi \partial_\n \phi \nonumber \\
&&  \left. - \frac{1}{2\Omega ^2} g^{\mu\nu} \partial_\m h \partial_\n h - \frac{V^{(0)}(\phi,h)}{\Omega ^4}\right\rbrace\,.
\label{eq:action1}
\ea
Following the notation of~\cite{Gialamas:2020snr} we will call this frame the ``intermediate frame" to account for the fact that, even though we have eliminated the nonminimal coupling that appears in the JF, we have not dealt with the quadratic terms yet.

%%%%%%%%%%%%%%%%%%%%%%%%%%%%%%%%%%%%%%%%%%%%%%
\section{Gildener-Weinberg approach}
\label{sec:GW_formalism}

Classically scale-invariant models containing multiple scalar fields are usually studied with the help of the Gildener-Weinberg formalism  \cite{Gildener:1976ih}\footnote{See also~\cite{Chang:2007ki, Foot:2007as, Foot:2007ay, Foot:2007iy, Holthausen:2009uc, Foot:2010av, AlexanderNunneley:2010nw, Foot:2010et, Foot:2011et, Lee:2012jn, Farzinnia:2013pga, Antipin:2013exa, Guo:2014bha, Radovcic:2014rea, Farzinnia:2014xia, Lindner:2014oea, Benic:2014aga, Kang:2014cia, Guo:2015lxa, Karam:2015jta, Ghorbani:2015xvz, Farzinnia:2015fka, Helmboldt:2016mpi, Hashino:2016rvx, Ahriche:2016cio, Khoze:2016zfi, Wu:2016jdo, Karam:2016rsz, Chataignier:2018kay, Loebbert:2018xsd, Brdar:2018vjq, YaserAyazi:2018lrv, Prokopec:2018tnq, Karam:2018mft, YaserAyazi:2019caf, Mohamadnejad:2019wqb, Kannike:2019upf, Jung:2019dog, Mohamadnejad:2019vzg, Lane:2019dbc, Kang:2020jeg, Dias:2020ryz, Braathen:2020vwo, Kannike:2020ppf, Kannike:2021iyh} for various applications of the formalism.}. In this approach, the perturbative minimization is realized at a definite energy scale due to the running of the coupling constants in the full quantum theory. Initially, one identifies the flat directions (FD) of the tree-level potential in the field space. These are directions along which the first derivatives of the potential with respect to each of the fields vanish. The flatness of the tree-level potential entails that the dynamics of the system is governed by the one-loop corrections which dominate along the FD. This way, the flatness is removed perturbatively and the physical vacuum of the theory is singled out from the valley of degenerate minima along the FD. In this section, we make use of the Gildener-Weinberg formalism and eventually end up with a single-field inflationary action.

\subsection{Tree-level minimization}

The tree-level potential after the Weyl rescaling of the JF action is given by
\beq
U^{(0)}(\phi,h) \equiv \frac{V^{(0)}(\phi,h)}{\Omega^4}=\frac{\left( \lambda_\phi \phi^4- \lambda_{h\phi} h^2 \phi^2 + \lambda_{h} h^4 \right)}{4 \left( \x _\phi \phi ^2 + \x _h h ^2 \right)^2}\,.
\label{U0_def}
\eeq
The first derivatives of $U^{(0)}(\phi,h)$ with respect to the two fields vanish along the trajectories in field space that satisfy the following conditions
\beq
\partial_{\phi} U^{(0)}(\phi,h) = 0\quad \Rightarrow \quad h^2 = \left( \frac{  \lambda _{h\phi}\, \xi _{\phi }+2 \lambda _{\phi }\, \xi _h }{\lambda _{h\phi} \,\xi _h +2 \lambda _h\, \xi _{\phi }} \right)\, \phi^2 \,,
\label{U0parphi}
\eeq
\beq
\partial_{h} U^{(0)}(\phi,h) = 0\quad \Rightarrow \quad \phi^2  = \left( \frac{  \lambda _{h\phi}\, \xi _h +2 \lambda _h\, \xi _{\phi } }{ \lambda _{h\phi}\, \xi _{\phi }+2  \lambda _{\phi }\, \xi _h}\right) \, h^2 \,.
\label{U0parh}
\eeq
A trajectory corresponds to a FD if it simultaneously satisfies both Eqs.~\eqref{U0parphi} and \eqref{U0parh}. Notice that, in our model, the two extremization conditions yield the same constraint and consequently they directly correspond to the FDs of $U^{(0)}(\phi,h)$. The two different signs correspond to the two independent FDs of the tree-level potential. We consider $\phi$ and $h$ to be positive definite and so, the relevant FD for our analysis is the one defined by the condition
\begin{equation}
   v_h =  \sqrt{ \frac{\lambda _{h\phi}\, \xi _{\phi }+2 \lambda _{\phi }\, \xi _h }{\lambda _{h\phi} \,\xi _h +2 \lambda _h\, \xi _{\phi }}}\, v_\phi\,, 
   \label{FD_cond}
\end{equation}
where the fields are at their VEV along the FD since it corresponds to the minimum of the potential. Note that for $\xi_h \ll 1$, if $v_\phi \sim M_{\rm P}$ and $\lambda_h \sim 0.1$, we find that the portal coupling needs to be extremely small, $\lambda_{h\phi} \sim 10^{-30}$. Upon employing Eq.~\eqref{FD_cond} we can compute the value of $U^{(0)}(\phi,h)$ along the FD in terms of the coupling constants of the model
\beq
U^{(0)}_{\rm min} \equiv U^{(0)}(v_\phi,v_h)=\frac{\left ( 4 \lambda_h \lambda_\phi - \lambda_{h \phi}^2 \right) M^4_{\rm P}}{16 \left[ \lambda_\phi \xi_h^2 + \xi_\phi \left ( \lambda_{h \phi} \xi_h+\lambda_h \xi_\phi \right ) \right]}\,.
\label{U0min}
\eeq
Notice that the minimum of the tree-level potential~\eqref{U0min} can be negative, zero or positive depending on the value of the combination $ 4 \lambda_h \lambda_\phi - \lambda_{h \phi}^2 $. On the contrary, had we applied the Gildener-Weinberg approach to the JF tree-level potential~\eqref{eq:scalar_pot} instead, the identification of the resulting extremization conditions would impose the constraint $\lambda_{h \phi}^2 = 4 \lambda_h \lambda_\phi$ and consequently, the minimum of~\eqref{eq:scalar_pot} would be fixed to zero. This freedom in specifying the minimum of the potential will play an important role in the next section where the one-loop corrections will be taken into account.

Having identified the FD of the tree-level potential we can move on to the computation of the mass matrix. Its elements are given by
\beq
M_{ij}^2 \equiv \frac{\partial^2 U^{(0)} }{\partial \Phi^i \, \partial \Phi^j}\biggr\rvert_{\Phi^i = v_{\Phi^i} , \Phi^j = v_{\Phi^j}}\,,
\eeq
where we denote $\left(\Phi^1,\Phi^2 \right)=\left(\phi,h \right)$ and $\u_{\Phi^i}$ are their respective VEVs. In terms of the ratio of the two VEVs we can define the mixing angle $\omega$ that corresponds to the angle between the $h=0$ axis in field space and the FD (see Fig.~\eqref{fig:Potentials}) as follows:
\beq
\omega \equiv \arctan{\left( \frac{v_h}{v_{\phi}} \right)}=\arctan{\left(\sqrt{ \frac{\lambda _{ \phi h}\, \xi _{\phi }+2 \lambda _{\phi }\, \xi _h }{\lambda _{ \phi h} \,\xi _h +2 \lambda _h\, \xi _{\phi }}} \right)} \,,
\label{eq:FD_angle_def}
\eeq
where in the last equality we have employed the condition \eqref{FD_cond}. We may now perform an orthogonal rotation described by the transformation
\beq
\left(\begin{array}{c}
\phi \\
h 
\end{array}
\right) = \begin{pmatrix} 
\cos{\omega} & -\sin{\omega} \\
\sin{\omega} & \cos{\omega}
\end{pmatrix} \left(\begin{array}{c}
s \\
\sigma
\end{array}
\right) \,,
\label{rotation}
\eeq
in order to move from the initial frame of fields $\left(\phi,h\right)$ to the \enquote{FD frame} $\left(s,\sigma \right)$ where the direction of the so-called \enquote{scalon} field $s$ is identified with the FD and $\sigma$ is the perpendicular direction.

Then, we may write the potential in terms of the FD frame fields in order to compute the mass matrix directly in that frame with $\left(\Phi^1,\Phi^2 \right)=\left(s,\sigma \right)$. The advantage of performing the rotation to the FD frame prior to the computation of the mass matrix is that the resultant matrix is diagonal. Thus, the mass eigenvalues for the fields $\left(s, \sigma \right)$ lie in the main diagonal and are given by the following expressions:
\ba
m^2_s &=&0\,,\\
m^2_{\sigma} &=& \frac{M_{\rm P}^4 \left( \lambda_{h \phi}\xi_h +2 \lambda_h \xi_\phi \right) \left( 2 \lambda_\phi \xi_h+\lambda_{h \phi}\xi_\phi \right)^2 \left[ \left( \lambda_{h \phi} +2 \lambda_\phi \right) \xi_h +\left( 2 \lambda_h +\lambda_{h \phi} \right) \xi_{\phi} \right]}{8\, v_h^2 \left[ \lambda_\phi \xi_h^2+\xi_\phi \left( \lambda_{h \phi}\xi_h+\lambda_h \xi_\phi \right) \right]^3} \,,
\ea
where we have once again employed Eq.~\eqref{FD_cond}.
As expected, the mass of $s$ is exactly zero at tree-level since it corresponds to the pseudo-Goldstone boson of broken classical scale symmetry. However, as we will see next, when quantum corrections are taken into account a nonzero mass will be generated for it. Furthermore, we identify the mass $m_\sigma$ with the measured value of the Higgs boson mass.

Along the FD ($\sigma =0$) the only relevant degree of freedom is the scalon $s$ which is related to $\phi$ and $h$ via
\ba 
s^2 = \phi^2 + h^2\,, \qquad s = \frac{\phi}{\cos \omega} = \frac{h}{\sin \omega}\,.
\label{phi_h_to_s}
\ea
The above relations can be easily verified by a simple inspection of the field space in Fig.~\eqref{fig:Potentials}. Upon employing Eqs.~\eqref{phi_h_to_s} we may rewrite the noncanonical kinetic terms for $h$ and $\phi$ in terms of $s$ as
\beq
 \frac{1}{\Omega ^2} \left[ \frac{1}{2} g^{\m\n} \partial_\m \phi \partial_\n \phi \nonumber +  \frac{1}{2} g^{\mu\nu} \partial_\m h \partial_\n h \right]= \frac{1}{\Omega ^2} \left[ \frac{1}{2} g^{\mu\nu} \partial_\m s \partial_\n s \right] \,,
\eeq
where, the nonminimal coupling functional expressed in terms of $s$ has the following form:
\be 
\frac{1}{\Omega^2} = \frac{1}{\xi_\phi \phi^2 + \xi_h h^2} = \frac{1}{\xi_s s^2} \,.
\ee
In the last equation, we have defined an \enquote{effective} nonminimal coupling constant for the scalon as
\beq
\xi_s \equiv \xi_\phi \cos^2 \omega + \xi_h \sin^2 \omega\,.
\label{xi_s_def}
\eeq
Finally, we perform the following field redefinition in order to render the kinetic term of $s$ canonical:
\be 
s_c - v_c = \int_{v_s}^s \frac{1}{\sqrt{\xi_s}} \frac{\dd s'}{s'} = \frac{1}{\sqrt{\xi_s}} \ln \frac{s}{v_s}\,.
\ee
The field $s_c$ is the one that drives inflation in our model and thus we shall refer to it as the inflaton field.

\subsection{One-loop effective potential}

The one-loop corrections along the flat direction for the canonical field $s_c$ at the scale $\Lambda$ may be written as
\be
\label{eq:one-loop-cor}
U^{(1)} (s_c) = \mathbb{A}\, s_c^4 + \mathbb{B}\, s_c^4 \ln \frac{s_c^2}{\Lambda^2} \,,
\ee
where in our model
\ba 
\mathbb{A} &=& \frac{1}{64 \pi^2 v_s^4} \left\lbrace M_h^4 \left( \ln \frac{M_h^2}{v^2_s} - \frac{3}{2} \right) + 6 M_W^4 \left( \ln \frac{M_W^2}{v^2_s} - \frac{5}{6} \right) + 3 M_Z^4 \left( \ln \frac{M_Z^2}{v^2_s} - \frac{5}{6} \right)  \right. \nonumber \\
 && \left. \qquad \ + 3 M_X^4 \left( \ln \frac{M_X^2}{v^2_s} - \frac{5}{6} \right) - 6 M_{N_R}^4 \left( \ln \frac{M_{N_R}^2}{v^2_s} - 1 \right) - 12 M_t^4 \left( \ln \frac{M_t^2}{v^2_s} - 1 \right)  \right\rbrace\,, \\ 
\mathbb{B} &=& \frac{\mathcal{M}^4}{64\pi^2 v^4_s} \,, \qquad \mathcal{M}^4 \equiv M_h^4 + 3 M_X^4 + 6 M_W^4 + 3 M_Z^4 - 6 M_{N_R}^4 - 12 M_t^4\,.
\ea
Minimizing~\eqref{eq:one-loop-cor}, we can determine the scale $\Lambda$ as
\be 
\Lambda = v_s \exp \left[ \frac{\mathbb{A}}{2 \mathbb{B}} + \frac{1}{4} \right]\,.
\ee
Then, we can express the one-loop correction as
\be 
U^{(1)} (s_c) = \frac{\mathcal{M}^4}{64\pi^2 v^4_s} s_c^4 \left[ \ln \frac{s_c^2}{v^2_s}  - \frac{1}{2} \right] \,.
\label{eq:one_loop}
\ee
One can see that the addition of the $U(1)_X$ gauge symmetry and in particular the mass of the extra gauge boson $X_\mu$ can render $\mathcal{M}^4$ positive if $3 M_X^4 - 6 M_{N_R}^4 \gtrsim \left( 317\ \rm GeV \right)^4$, which in turn implies that the one-loop potential is bounded from below at large field values. From the one-loop corrections we can obtain the radiatively generated mass for the $s$ scalar
\be 
m^2_s = \frac{\mathcal{M}^4}{8\pi^2 v^2_s}\,.
\label{eq:m_s}
\ee
Notice that at the minimum, the one-loop correction~\eqref{eq:one_loop} is negative. With this observation, the choice to consider the one-loop corrections in the ``intermediate frame"~\eqref{eq:action1}, and not in the JF action~\eqref{eq:action_jf} is justified. Had we opted for the latter, the extremization conditions for the tree-level JF potential would fix its value to zero along the flat direction, as we have already mentioned, and thus the full one-loop effective potential (tree-level + one-loop) would correspond to an anti-de Sitter vacuum. This issue can of course be easily circumvented by including a positive cosmological constant in the effective potential in order to reach a Minkowski vacuum, albeit in this case, the model ceases to be scale invariant and instead is characterized as quasiscale invariant. 

We now require that the full one-loop effective potential is zero at $v_s$ which can be realized once we assume that $ 4 \lambda_h \lambda_\phi - \lambda_{h \phi}^2>0$, so that $U^{(0)}_{\rm min}>0$. Then we may write
\be 
\label{eq:CC}
U_{\rm eff}(v_s) = U^{(0)}_{\rm min} + U^{(1)} (v_s) = 0\,,
\ee
which finally yields
\be 
U_{\rm eff} (s_c) = \frac{\mathcal{M}^4}{128 \pi^2} \left[ \frac{s^4_c}{v^4_s} \left( 2 \ln \frac{s^2_c}{v^2_s} - 1 \right)  + 1  \right]\,.
\label{eq:Veff}
\ee
Note that the condition~\eqref{eq:CC} effectively means that the cosmological constant can potentially be generated from two or higher-order loop corrections.

The VEV of the inflaton $s_c$ is associated with the reduced Planck mass via the value of the effective nonminimal coupling constant~\eqref{xi_s_def} as
\be
\label{eq:xi_svev}
v^2_s = \frac{M^2_{\rm P}}{\xi_s}\,,
\ee
and thus, it is evident that in principle $v_s$ can be super-Planckian for $\xi_s<1$. Indeed, as we will see in Sec.~\ref{sec:Inflation}, this is exactly the case in our model since observationally viable inflation requires $\xi_s \lesssim \mathcal{O}(10^{-3})$.

Finally, the effective action along the FD written explicitly in terms of the inflaton field reads
\beq
S=\int{\dd^4 x \sqrt{-g}}\left\lbrace \frac{1}{2} \left[ g^{\mu\nu} R_{\mu\nu} + \alpha R^2 + \beta R_{\mu\nu} R^{\mu\nu}  \right]-\frac{1}{2} g^{\mu\nu} \partial_{\mu} s_c \partial_{\nu}s_c - U_{\rm eff}(s_c) \right\rbrace\,.
\label{JF_FD_action}
\eeq
In the next section, our objective is to identify and employ the appropriate transformations in order to remove the higher-curvature terms and eventually recast the effective action~\eqref{JF_FD_action} in the EF with the gravity sector consisting solely of the Einstein-Hilbert term.

%%%%%%%%%%%%%%%%%%%%%%%%%%%%%%%%%%%%%%%%%%%%%%
\section{Einstein frame representation}
\label{sec:EF_rep}

%In this section, in order to obtain the predictions of the model for the cosmological observables, we will pass from the ``intermediate frame" of Eq.~\eqref{eq:action1} or~\eqref{JF_FD_action} into the EF applying the procedure which was also described in~\cite{Afonso:2017bxr,Annala2020}.

In this section, in order to obtain the predictions of the model for the cosmological observables, we will pass from the ``intermediate frame" of Eq.~\eqref{eq:action1} or~\eqref{JF_FD_action} into the EF applying the procedure which was outlined in~\cite{Annala2020} (see also~\cite{Afonso:2017bxr}).

\subsection{The Legendre transformation}

The action~\eqref{JF_FD_action} can be cast in the form
% \footnote{From now on and for reasons which will be discussed later, we will use only the symmetric Ricci tensor $R_{(\m\n)}$ and in order to speed up notation we discard the parentheses.}
\be
S =  \int \dd^4 x \sqrt{-g}  \left[ \frac{1}{2}C(g_{\m\n},R_{\m\n}) + \mathcal{L}_m
(g_{\m\n},\,s_c,\, \partial_\mu s_c) \right]\,,
\label{eq:action2}
\ee
where we have defined the ``curvature" function
\be
C(g_{\m\n},R_{\m\n})=g^{\m\n} R_{\m\n} + \alpha R^2 + \beta R_{\m\n} R^{\m\n}\,,
\label{eq:curvfun}
\ee
and the matter Lagrangian density
\be
\mathcal{L}_m (g_{\m\n},\,s_c,\, \partial_\mu s_c)=  -\frac{1}{2} g^{\m\n} \partial_\m s_c \partial_\n s_c - U_{\rm eff}(s_c)\,.
\label{eq:Lagmat}
\ee
From this point onward the $\partial_\m s_c$ dependence in the argument of $\mathcal{L}_m$ will be ignored for brevity. Now, upon introducing the auxiliary field $\Sigma_{\m\n}$ the action becomes
\be
S =  \int \dd^4 x \sqrt{-g}  \left[ \frac{1}{2}C(g_{\m\n},\Sigma_{\m\n},s_c) + \frac{1}{2} \frac{\partial C}{\partial \Sigma_{\m\n}} \left(R_{\m\n}-\Sigma_{\m\n} \right)
+\mathcal{L}_m (g_{\m\n},s_c) \right]\,.
\label{eq:action3}
\ee
It is trivial to see that the variation $\d S /\d \Sigma_{\m\n} =0$ gives that $\Sigma_{\m\n} =R_{\m\n}$. The advantage of action~\eqref{eq:action3} is that it is linear in the Ricci tensor so it is one step closer to the final EF action. We introduce the new variable $q^{\m\n}$ which is defined as
\be
\sqrt{-q} q^{\m\n} = \sqrt{-g} \frac{\partial C}{\partial \Sigma_{\m\n}}\,, 
\label{eq:qmn}
\ee
where $q=\det(q_{\m\n})$ and $q^{\m\n} q_{\m\l}=\d^\n_{\spc\l}$. Using~\eqref{eq:qmn} we can solve the  $\Sigma_{\m\n}$ in terms of the $g_{\m\n}$, $s_c$ and $q_{\m\n}$, thus the action can be written as
\beq
S =  \int \dd^4 x \left\lbrace \frac{\sqrt{-q}}{2}  q^{\m\n} R_{\m\n} -\frac{\sqrt{-g}}{2} \left[ \frac{\partial C}{\partial \Sigma_{\m\n}} \Sigma_{\m\n}(q_{\m\n}, g_{\m\n}, s_c) - C(q_{\m\n}, g_{\m\n}, s_c) -2\mathcal{L}_m (g_{\m\n},s_c) \right] \right\rbrace\,. 
\label{eq:action4}
\eeq
The gravitational sector of~\eqref{eq:action4} is the typical Einstein-Hilbert term for the metric $q_{\m\n}$. Varying the action~\eqref{eq:action4} with respect to $g_{\m\n}$ (see Appendix~\ref{sec:appendixA}) will give us $g_{\m\n}$ as a function of $q_{\m\n}$, $s_c$ and $\partial_\m s_c$. This way we obtain that
\ba
\frac{1}{\sqrt{-g}}\frac{\d S}{\d g^{\m\n}} = && -\frac{1}{4(\b+4\a)}\frac{\sqrt{-q}}{\sqrt{-g}}q^{\s\l}g_{\s\m}g_{\l\n} \nonumber
\\&& +\frac{1}{4\b}\frac{q}{g}\left(q^{\s\l}q^{\r\d}g_{\l\d}g_{\r\n}g_{\s\m} -\frac{\a}{\b+4\a}q^{\d\r}g_{\d\r}q^{\s\l}g_{\s\m}g_{\l\n} \right) \nonumber
\\&& +\frac{1}{2}g_{\m\n} \left[\frac{1}{\b+4\a}\left( \frac{1}{2}+\frac{\a}{8\b}\frac{q}{g}q^{\l\s}g_{\l\s}q^{\r\d}g_{\r\d}\right) -\frac{q}{g}\frac{1}{8\b}q^{\l\s}q^{\d\r}g_{\l\d}g_{\s\r}\right] \nonumber
\\ && + \frac{1}{2}g_{\m\n} \left(\frac{1}{2}g^{\l\s} \partial_\l s_c \partial_\s s_c + U_{\rm eff}(s_c) \right) - \frac{1}{2}\partial_\m s_c \partial_\n s_c =0\,.
\label{eq:daction}
\ea
which will help us to solve the metric $g_{\m\n}$ in terms of the metric $q_{\m\n}$ and the inflaton field\footnote{Equation~~\eqref{eq:daction} has been also derived in~\cite{Annala2020}, with a missing $1/2$ factor in the parenthesis in the third line. We think that this is only a misprint as our final results are in absolutely agreement with these of~\cite{Annala2020}.}.
\subsection{The disformal transformation}

Another useful type of metric transformation is the disformal transformation~\cite{Bekenstein:1992pj, Zumalacarregui:2013pma, Minamitsuji:2014waa, Tsujikawa:2014uza, Domenech:2015hka}, a generalization of the well-known conformal transformation. It can be used in order to bring complicated actions, e.g.~\eqref{JF_FD_action}, into the EF. This is of the form
\be
g_{\m\n}=A\, q_{\m\n} +B\, \partial_\m s_c \partial_\n s_c \,,
\label{eq:disfgmn}
\ee
where the coefficients $A$ and $B$ are functions of $s_c$ and $X_q$ with
\be
X_q \equiv -\frac{1}{2}q^{\m\n}\partial_\m s_c \partial_\n s_c\,.
\label{eq:kinterms}
\ee
The relation that correlates the determinants of the metrics $g_{\m\n}$ and $q_{\m\n}$ can be easily obtained upon substituting the general form of the disformal transformation~\eqref{eq:disfgmn} into $\det\left( q^{\m\s} g_{\m\n}\right)= q^{-1} g$. That is,
\be
g=qA^3\left(A-2B X_q \right)\,.
\label{eq:determinants}
\ee
For our computation we also need the inverse metric $g^{\m\n}$. Following~\cite{Firouzjahi:2018xob} we obtain that
\be
g^{\m\n}=\bar{A}\, q^{\m\n} +\bar{B}q^{\m\l} q^{\n\s} \partial_\l s_c \partial_\s s_c\,,
\label{eq:disfgmn_inv}
\ee
where
\beq
\bar{A}=\frac{1}{A}\,,\quad \bar{B}= -\frac{B}{A^2 -2ABX_q} \,.
\label{eq:invcofs}
\eeq
Finally, using~\eqref{eq:kinterms} and~\eqref{eq:disfgmn_inv} it is quite trivial to prove that the kinetic terms for the metric $g_{\m\n}$ can be expressed in terms of the kinetic terms for the metric $q_{\m\n}$ as
\beq
X_g = \bar{A} X_q -2\bar{B} {X_q}^2\,.
\label{eq:disfkinet}
\eeq
Now, we can substitute~\eqref{eq:disfgmn} and~\eqref{eq:disfkinet} in~\eqref{eq:daction}. This substitution will give us two algebraic equations. Each equation arises from the requirement that the coefficients of $q_{\m\n}$, and $\partial_\m s_c \partial_\n s_c $ must vanish identically. These equations are listed below:
\ba
 \frac{1}{16 \b   (4 \a +\b) R_5}&&\left( 4(4 \a +\b ) A^2 -4  \b  A\sqrt{R_5}-4 \a  A R_2- (4 \a +\b ) R_3+4 \b R_5+\a R_2^2 \right) \nonumber
\\ && +\frac{U_{\rm eff}(s_c)}{2} - \frac{X_g}{2}=0\,,
\label{eq:eq1}
\ea
\ba
 \frac{1}{16 \b   (4 \a +\b) R_5}&&\Big(4  (4 \a +\b )R_4 -4  \b R_1 \sqrt{R_5}-4 \a  R_2 R_1- (4 \a +\b )B R_3 + 4 \b B R_5 \nonumber
\\ &&  +\a B R_2^2 \Big) + \frac{B U_{\rm eff}(s_c)}{2}-\frac{B X_g}{2}-\frac{1}{2}=0\,,
\label{eq:eq2}
\ea
where the functions $R_i$ are given in Appendix~\ref{sec:appendixB}. Equations~\eqref{eq:eq1} and~\eqref{eq:eq2} accorded well with Eqs.~(6.44) and~(6.45) of~\cite{Annala2020}. Our aim is to solve the system~\eqref{eq:eq1} and~\eqref{eq:eq2}, but this a very difficult task. However, we can approximate the solutions assuming that in the slow roll approximation the higher-order kinetic terms are negligible at least during inflation~\cite{Tenkanen:2020cvw}, but also during reheating~\cite{Karam:2021sno}.  Thus the approximate solution is of the form
\ba
A &=& a_0 +a_1  X_q +\mathcal{O}(X_q^2) \,,  \nonumber \\
B &=& b_0 +b_1  X_q +\mathcal{O}(X_q^2)\,.
\label{eq:solslowroll}
\ea
By substituting~\eqref{eq:solslowroll} in the system~\eqref{eq:eq1} and~\eqref{eq:eq2} and expanding in terms of the kinetic term~\eqref{eq:kinterms}, we can solve for the coefficients $a_i$, and $b_i$ after forcing that the coefficient of each order vanishes identically. These coefficients are listed in Appendix~\ref{sec:appendixB}.

Having done all the groundwork, we can substitute the solution~\eqref{eq:solslowroll} with the coefficients~\eqref{eq:ai} to the matter sector~\eqref{eq:lagG} and expand again in the kinetic term. This gives us the final EF action
\ba
S =  \int \dd^4 x \sqrt{-q}  && \left[ \frac{1}{2}   q^{\m\n} R_{\m\n} + K(s_c) X_q -\ubar(s_c)  +\mathcal{O}(X_q^2)\right]\,, 
\label{eq:final_action}
\ea
with 
\ba
K(s_c) = \frac{1}{1+\ta\, U_{\rm eff}(s_c)}\spc \text{and} \spc \ubar(s_c)= \frac{U_{\rm eff}(s_c)}{1+\ta\, U_{\rm eff}(s_c)}\,, 
\label{eq:KUbar}
\ea
where we have defined the \enquote{effective} higher-curvature coupling $\ta \equiv 2\b+8\a$. To avoid ghosts we require that $K > 0$ and thus $\ta>0$. This is true if both $\a$ and $\b$ are positive, but also if $\b>-4\a$. Regarding the magnitude of the parameter $\ta$, according to~\cite{Karam:2021sno}, unitarity considerations suggest that $\ta \lesssim 10^{21}$.

\begin{figure}[h!]
\centering
\includegraphics[width=0.7\linewidth]{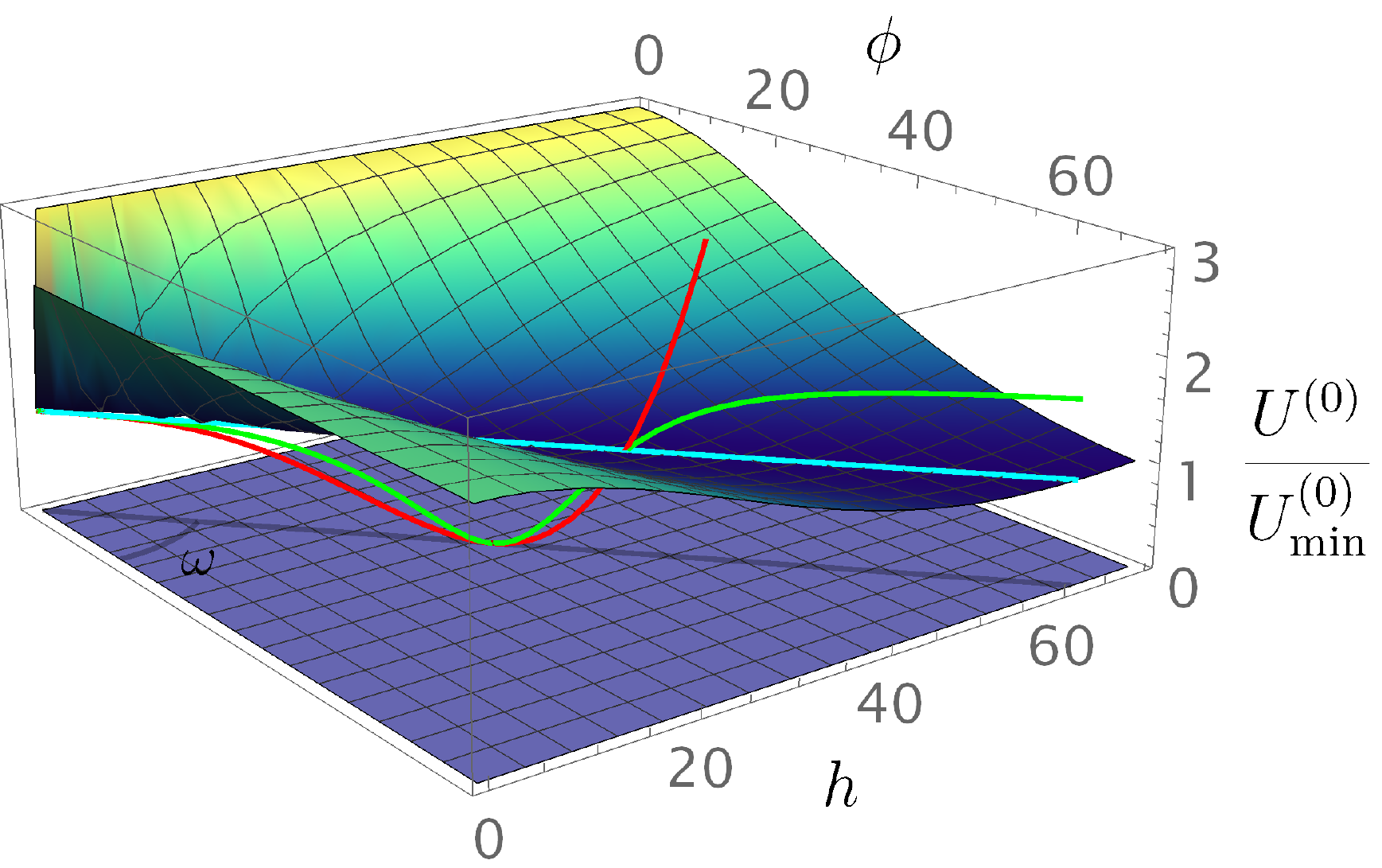}
\caption{\sf The normalized tree-level potential $U^{(0)}(\phi,h)/U^{(0)}_{\rm min}$~\eqref{U0_def} and its flat direction (cyan line). We also plot the normalized one-loop corrected potential along the flat direction $U_{\rm eff} (s_c)/U_{\rm eff} (0)$~\eqref{eq:Veff} (red curve) and the normalized inflaton potential $\bar{U}(s_c)/\bar{U}(0)$~\eqref{eq:KUbar} (green curve). The values of the parameters are $\ta =10^9 ,\,\xi_s=10^{-3} ,\,$ and $\mathcal{M} \simeq 0.0357$. For illustrative purposes we have chosen the values of the couplings $\lambda_\phi,\,\lambda_h,\,\lambda_{h\phi}$ such that the mixing angle \eqref{eq:FD_angle_def} has the unrealistic value of $\omega \simeq 0.732$.
}\label{fig:Potentials} 
\end{figure}

We have thus far mentioned various potentials and in order to demonstrate their qualitative differences we plot them collectively in Fig.~\ref{fig:Potentials}. The surface with the color gradient corresponds to the normalized two-field tree-level JF potential $U^{(0)}(\phi,h)/U^{(0)}_{\rm min}$ as given in Eq.~\eqref{U0_def}. Its FD which we have identified by means of the GW approach is depicted with the cyan line. Once quantum corrections are taken into account, the one-loop corrected potential \eqref{eq:Veff} with a unique minimum singled out from the valley of degenerate vacua along the FD is obtained and we depict it with the red curve in its normalized form $U_{\rm eff} (s_c)/U_{\rm eff} (0)$. Finally, the normalized inflationary potential $\bar{U}(s_c)/\bar{U}(0)$ for our model~\eqref{eq:KUbar} is depicted with the green curve. Notice that $\ubar(s_c)$ exhibits plateaus on both sides of the minimum and thus it is suitable for both small field inflation and large field inflation i.e. excursions of the inflaton field in the regions $s_c < v_s$ and $s_c > v_s$ respectively.

In the end, having started with a general scale invariant action which involves adimensional matter-gravity and matter-matter couplings, we have obtained an action with a noncanonical scalar field that is minimally coupled to the usual Einstein-Hilbert action at the expense of negligible higher-order kinetic terms and a modified  potential, which as we will see next is suitable for successful inflation in accordance with observations.

%%%%%%%%%%%%%%%%%%%%%%%%%%%%%%%%%%%%%%%%%%%%%%
\section{Slow roll approximation and contact with observations}
\label{sec:Inflation}

In this section, in order to constrain the parametric space of our model, we compare its predictions for the cosmological observables with their corresponding latest observational bounds as set by the Planck collaboration.

\subsection{Inflationary observables and number of $e$-folds}

The number of $e$-folds elapsed during the inflationary phase can be obtained in terms of the potential $\ubar(s_c)$ and the kinetic term coupling function $K(s_c)$ as
\beq
N_\star \equiv N(s_{c\star}) = \int_{s_{c,\rm end}}^{s_{c\star}}{K(s_c)\frac{\ubar(s_c)}{\ubar'(s_c)}}\dd s_c\,,
\eeq
where primes are used to denote differentiation with respect to the argument, while the subscripts $``\star"$ and $``{\rm end}"$ denote quantities at the time of horizon crossing of the pivot scale $k_\star$, and at the end of inflation respectively. The potential slow roll parameters (PSRPs) are defined as
\beq
\epsilon(s_c)\equiv \frac{1}{2 K(s_c)}\left[\frac{\ubar'(s_c)}{\ubar(s_c)}\right]^2\,, \quad \eta(s_c)\equiv \frac{1}{\ubar(s_c) \sqrt{K(s_c)}}  \left[\frac{\ubar'(s_c)}{\sqrt{K(s_c)}} \right]'\,.
\eeq
During slow roll inflation $\epsilon(s_c) \ll 1$ and $|\eta(s_c)| \ll 1$ and inflation ends, to a very good approximation, when $\epsilon(s_c) = 1$. The values of the cosmological observables in the slow roll approximation can be obtained in terms of the PSRPs evaluated at the time of horizon crossing $\eps_\star \equiv \eps(s_{c\star})$ and $\eta_\star \equiv \eta(s_{c\star})$. The observables that are relevant for our analysis are the tensor-to-scalar ratio
\beq
r \simeq 16\, \eps_\star\,,
\eeq
the tilt of the scalar power spectrum
\beq
n_s \simeq 1-6 \,\eps_\star+2\, \eta_\star \,,
\eeq
and the amplitude of scalar perturbations
\beq
A_s \simeq \frac{1}{24 \pi^2} \frac{\ubar(s_{c\star})}{\eps_\star}\,.
\eeq
The Planck collaboration~\cite{Akrami:2018odb} has set the following bounds on the values of the observables:
\begin{equation}\label{eq:Planck_constr}
    A_s\simeq2.1\times10^{-9},\qquad\quad n_s=\left\{\begin{matrix}\left(0.9607,0.9691\right),&\ 1\sigma\text{ region}\\\left(0.9565,0.9733\right),&\ 2\sigma\text{ region}\end{matrix}\right.,\qquad\quad r\lesssim 0.056\,.
\end{equation}
The number of $e$-folds at the pivot scale $k_\star$ assuming instantaneous reheating can be very well approximated as~\cite{Akrami:2018odb,Liddle:2003as}
\be
 N_\star  =  \ln \left[ \left(\frac{\pi^2}{30} \right)^{\frac{1}{4}} \frac{( g_{s,0})^{ \frac{1}{3}}}{ \sqrt{3} } \frac{T_0}{H_0} \right]
 - \ln\left[ \frac{k_\star}{  a_0 H_0} \right] 
 + \frac{1}{4} \ln\left[\frac{\ubar^2(s_{c\star})}{\rho_{\rm end}} \right] - \frac{1 }{12}  \ln  \left[ g_{s,{ \rm reh}} \right]  
\, ,
  \label{eq:Nstar1}  
 \ee   
where the subscripts $ ``0" $ and $``{\rm reh}"$ denote quantities at the present epoch and reheating phase  respectively. With $\rho$ we denote the energy density. The entropy density degrees of freedom $g_s$ have the values $g_{s,0}=43/11$ and $ g_{s,{ \rm reh}} = \mathcal{O}(100)$ in our model and for reheating temperatures $\sim 1\,\mathrm{TeV}$ or higher. At the present epoch the CMB temperature and the Hubble constant are $T_0=2.725\,\mathrm{K}$ and $H_0=67.6\,\mathrm{km/s/Mpc}$ respectively and we fix the pivot scale to $k_\star = 0.002\, \text{Mpc}^{-1}$.
Being more accurate we should calculate $\rho_{\rm end}$ by taking into account that the Hubble slow roll parameter $\eps_1 \equiv -\Dot{H}/H^2$ is exactly $\eps_1=1$ at the end of inflation. This condition gives that $\rho_{\rm end} =3\bar{U} (s_{c,\rm end})/2 $. Using this and writing~\eqref{eq:Nstar1} in terms of the potential~\eqref{eq:Veff} we can make explicit the dependence of the number of $e$-folds on the parameter $\ta$, that is,
\be
N_\star = 64.3+\frac{1}{4} \ln \left(  \frac{2 {U_{\rm eff}^{\star}}^2 (1+\ta U_{\rm eff}^{\rm end})}{3  U_{\rm eff}^{\rm end} (1+\ta U_{\rm eff}^\star)^2}\right)\,.
\label{eq:Nstar2}
\ee  
In \cite{Gialamas:2019nly,Gialamas:2020snr}, the higher-order kinetic terms appearing in the action~\eqref{eq:final_action} have been taken into account in the calculation of $\rho_{\rm end}$, but as it is shown there, only an insignificant correction arises in the numerical factor of the number of $e$-folds. In addition, in~\cite{Gialamas:2019nly, Lykkas:2021vax} the reheating mechanism in $R^2$ Palatini inflationary models has been studied, but beyond the case of instantaneous reheating, allowing a wider range for the number of $e$-folds for various values of the equation of state parameter.

\subsection{Small and large field inflation}

Prior to performing the full parametric space investigation for the inflationary predictions of the model, we mention some asymptotic limits with respect to the value of the effective nonminimal coupling $\xi_s$. For $\xi_s \ll 1$ and $\ta = 0$ we find that the predictions for both small field inflation (SFI) and large field inflation (LFI) correspond to those of quadratic inflation,
\be 
n_s \simeq 1 - \frac{2}{N_\star} \,, \qquad r_{0} \simeq \frac{8}{N_\star}\,,
\ee
where $r_0$ denotes the tensor-to-scalar ratio for $\ta = 0$. On the other hand, for $\xi_s \gg 1$ we find
\be 
n_s \simeq 1 - \frac{3}{N_\star}\,,
\ee
for both SFI and LFI, while 
\be 
r_0 \simeq \frac{16}{N_\star}\,\quad \text{(for LFI)}\,,\qquad r_0 \simeq 0\,\quad \text{(for SFI)\,,}
\ee
Note that the first limit corresponds to the prediction of quartic inflation. When $\ta \neq 0$, the predictions for $n_s$ remain the same but $r$ gets modified as~\cite{Enckell:2018hmo, Gialamas:2020snr}
\be 
\label{eq:r}
r = \frac{r_0}{1 + \ta U^\star_{\rm eff}} = \frac{r_0}{1 + \frac{3}{2} \pi^2 \ta A_s r_0 }\,,
\ee
therefore, the presence of the parameter $\ta$ results in a suppression of the value of tensor-to-scalar ratio.

\begin{figure}[h!]
\includegraphics[width=0.5\linewidth]{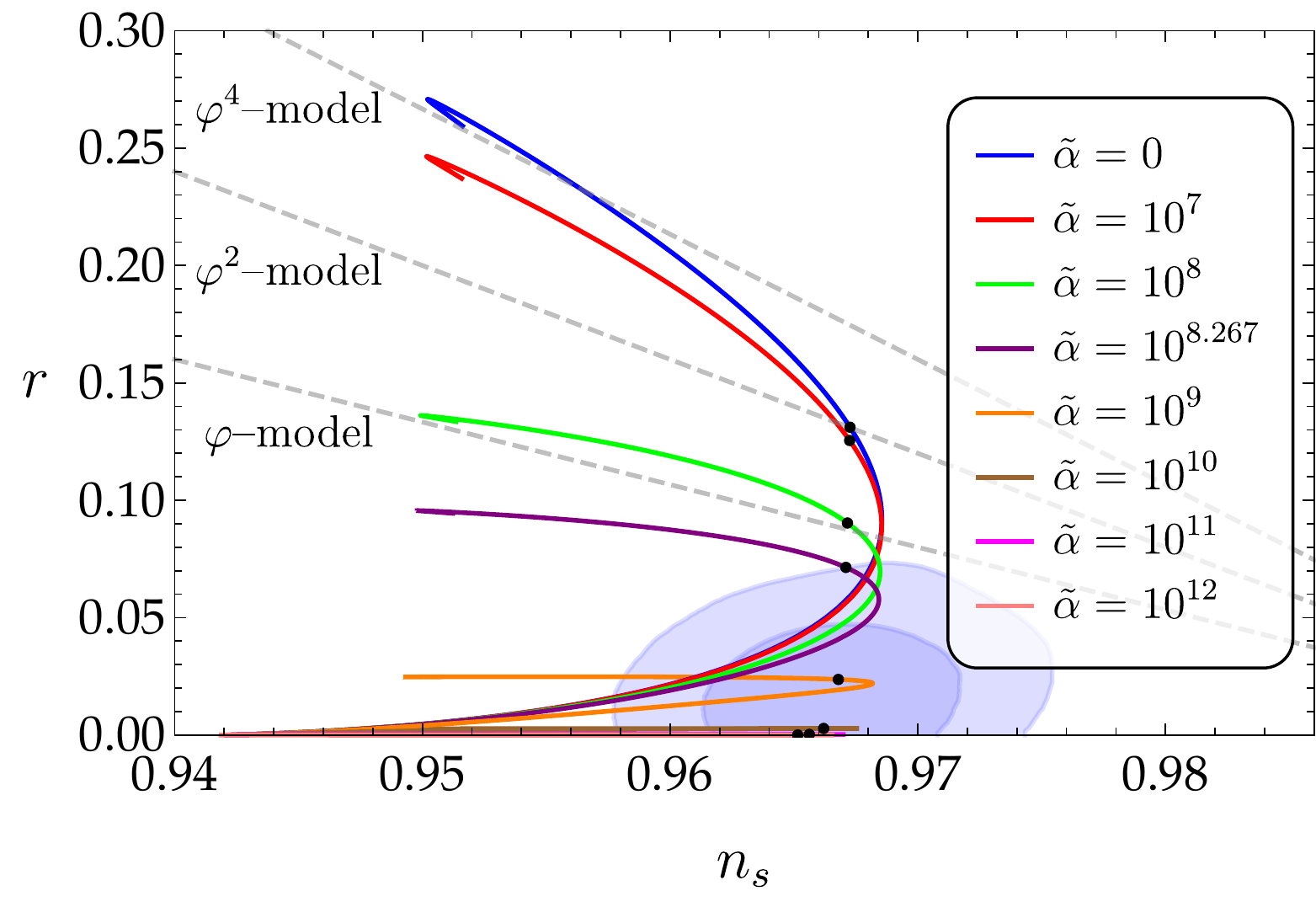}
\includegraphics[width=0.5\linewidth]{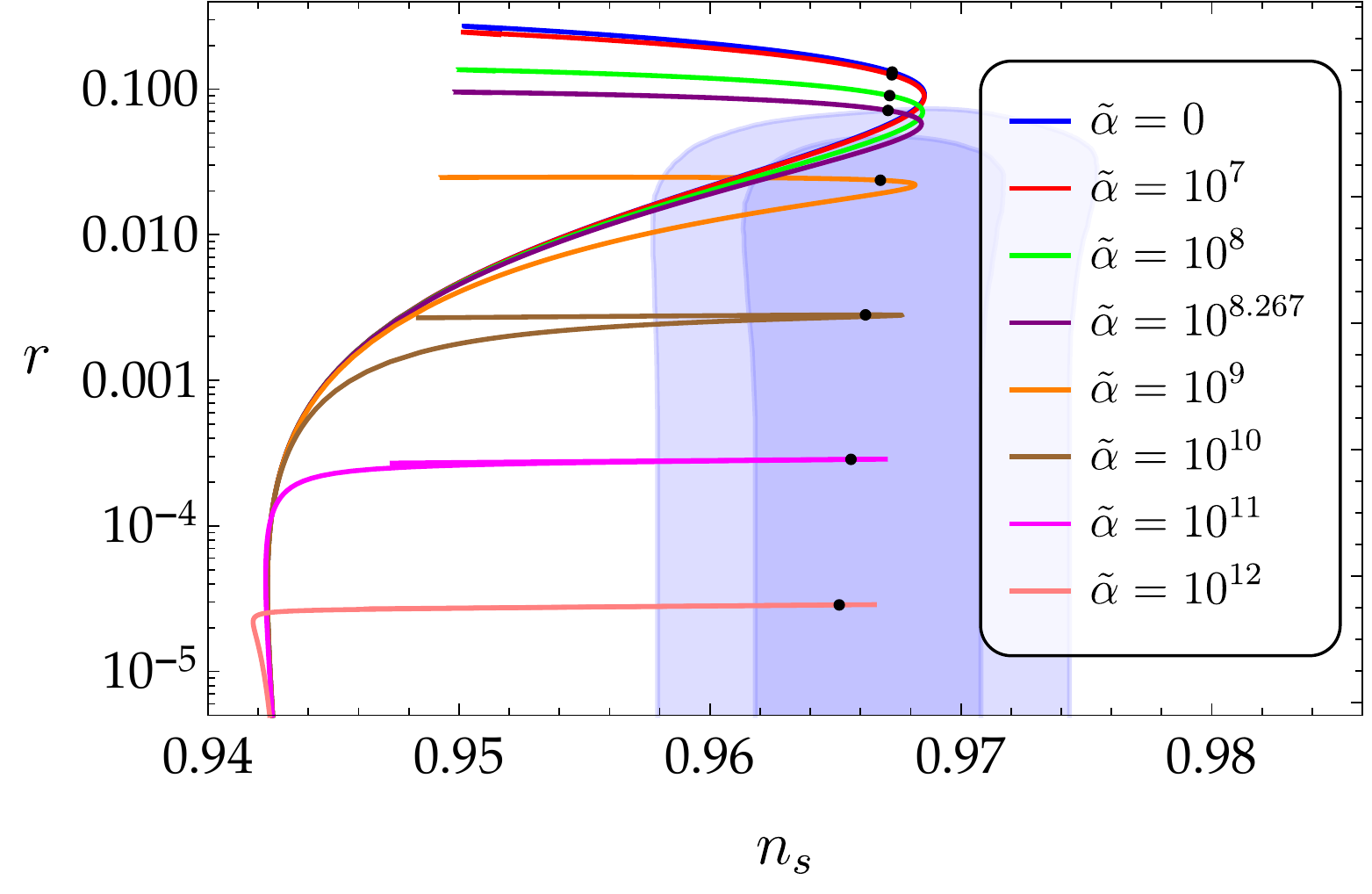}
\caption{\sf The predictions for the tensor-to-scalar ratio ($r$) and the tilt of the scalar spectrum ($n_{\rm s}$) as $\xi_{\rm s}$ ranges from $\xi_{\rm s} \ll 1$ to $\xi_{\rm s} \gg 1$ for various values of $\ta$. For each one of the curves, the black dot corresponds to the $\xi_{\rm s} \rightarrow 0$ limit (see Table~\ref{table:xi_s_zero_limit}) and $\xi_{\rm s}$ increases monotonically as we move away from it along each one of the two directions on the curve. The upper (lower) part of a curve with respect to its $\xi_{\rm s} \rightarrow 0$ limit, corresponds to the predictions of large (small) field inflation. On the left (right) panel, the predictions are depicted in linear (logarithmic) scale.}
\label{fig:r_ns}
\end{figure}

Let us now turn to the full analysis of the parametric space of our model with respect to its predictions for the cosmological observables. For each given set of values for the parameters $\ta$ and $\xi_{\rm s}$, we have employed Eq.~\eqref{eq:Nstar2} to obtain the number of $e$-folds that complies with the constraints from reheating, while the value of $\mathcal{M}$ has been fixed in each case such that we always have $A_s =2.1 \times 10^{-9}$ at $k_\star = 0.05\, \text{Mpc}^{-1}$ in accordance with the bounds set by the Planck collaboration. For both SFI and LFI, we have considered various values of $\ta$ and a wide range of values for $\xi_{\rm s}$ ranging from $\xi_{\rm s} \ll 1$ to $\xi_{\rm s} \gg 1$ and in Fig.~\ref{fig:r_ns} we plot the corresponding predictions for the tensor-to-scalar ratio and the scalar tilt against the $68\%$ (dark blue) and $95\%$ (light blue) CL regions for $n_{\rm s}$ and $r$ at $k_* =0.002\, \text{Mpc}^{-1}$ as obtained with the combined data from Planck+BK15+BAO~\cite{Akrami:2018odb}.

The different curves correspond to fixed values of $\ta$, while $\xi_{\rm s}$ ranges along the curves with the black dot on each curve corresponding to the $\xi_{\rm s} \rightarrow 0$ limit. These dots also designate the transition point between the predictions of SFI and LFI with the lower (upper) part of each curve corresponding to small (large) field inflation. Evidently in the limit of small $\xi_{\rm s}$ the predictions of SFI and LFI are identical. As we move away from the $\xi_{\rm s} \rightarrow 0$ limit along a given curve in both directions $\xi_{\rm s}$ increases monotonically with the top end of the curves corresponding to $\xi_{\rm s}$ values of $\mathcal{O}\left( 10^{8} \right)$ and the bottom end (more clearly shown in the right panel of Fig.~\ref{fig:r_ns}) to values of $ \mathcal{O}\left( 10^{-1} \right)$.

\begin{table}[h!]
\center{
\begin{tabular}{|c|c|c|c|c|c|c|c|c|}
 \hline\hline
$\ta $   &  $0$ & $10^{7}$ & $10^{8}$ & $1.85 \times 10^{8}$ & $10^{9}$ & $10^{10}$ & $10^{11}$ & $10^{12}$ \\
 \hline
  $r$ & 0.13090 & 0.12526 & 0.09022 & 0.07134 & 0.02368 & 0.00282 & 0.00029 & 0.00003  \\
 \hline
  $n_s$ & 0.96727 & 0.96726 & 0.96717 & 0.96711 & 0.96681 & 0.96621  & 0.96563 & 0.96517  \\
 \hline
  $N_\star$ & 60.6 & 60.6 & 60.4 & 60.3 & 59.8 & 58.8 & 58.0 & 57.3 
   \\
 \hline\hline
\end{tabular}
\caption{\sf The predicted values for the tensor-to-scalar ratio ($r$), tilt of the scalar spectrum ($n_{\rm s}$) and number of efolds ($N_\star$), in the limit $\xi_{\rm s} \rightarrow 0$ for various values of $\ta$.}
\label{table:xi_s_zero_limit}}
\end{table}

\begin{figure}[h!]
\centering
\includegraphics[width=0.7\linewidth]{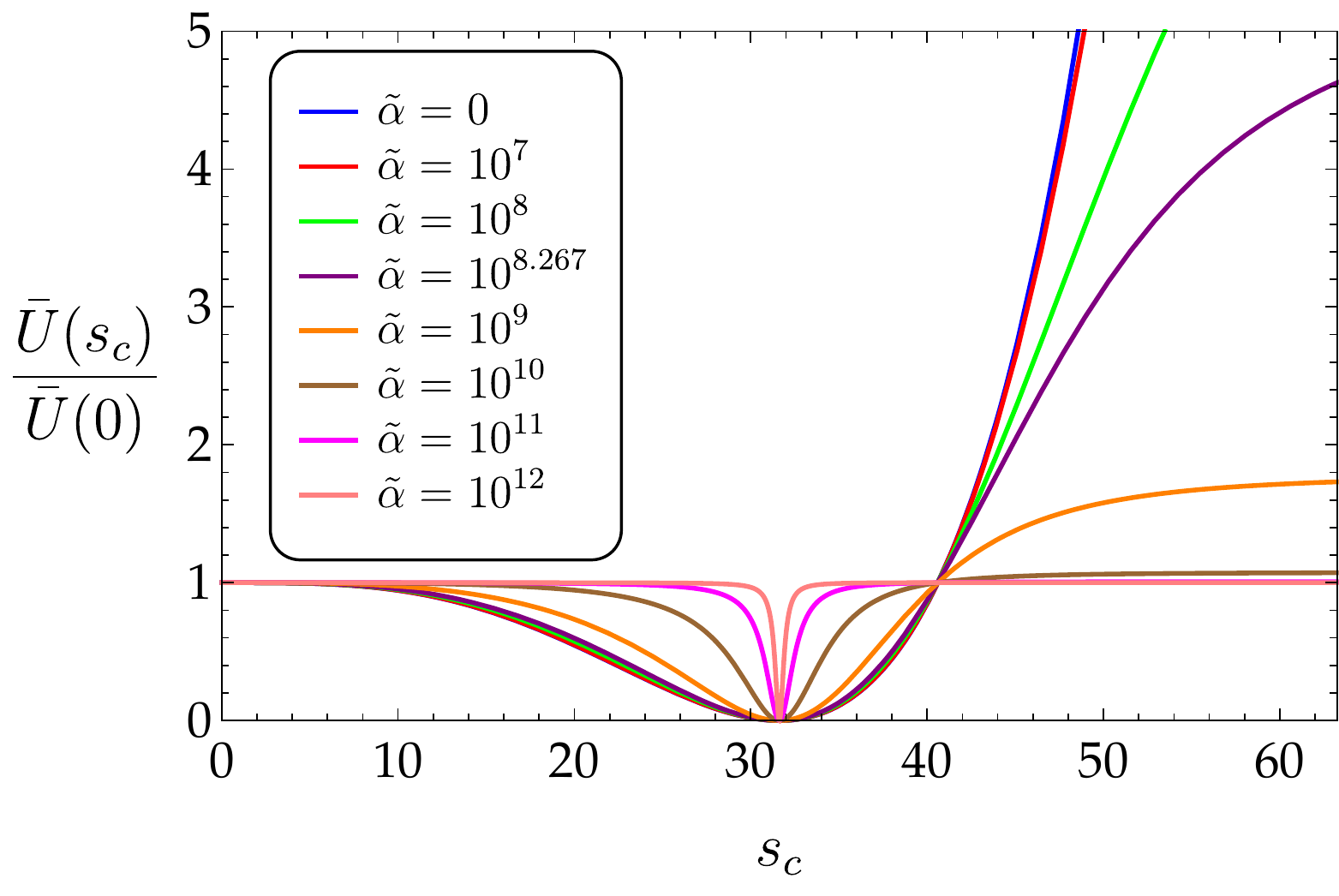}
\caption{\sf The normalized potential $\ubar(s_c)$ for $\xi_{\rm s}=0.001$ and various values of $\ta$. In the limit $\ta \gg 1$, the potential becomes symmetric about its VEV and consequently the predictions for small and large field inflation are identical.}
\label{fig:Ubar_vs_a_tilde}
\end{figure}

The effect of $\ta$ on the inflationary predictions is to suppresses the value of $r$ (cf. Eq.~\eqref{eq:r}). This effect becomes important for values $\ta \gtrsim 10^{6}-10^{7}$. As the right panel of Fig.~\ref{fig:r_ns} reveals, for sufficiently large values of $\ta$ the predictions of SFI and LFI are identical along an extended range of values of $\xi_{\rm s}$. This can be understood via the shape of the inflationary potential that becomes symmetric about the location of the VEV for $\ta \gg 1$, see Fig.~\ref{fig:Ubar_vs_a_tilde}.

\begin{figure}[h!]
\centering
\includegraphics[width=0.7\linewidth]{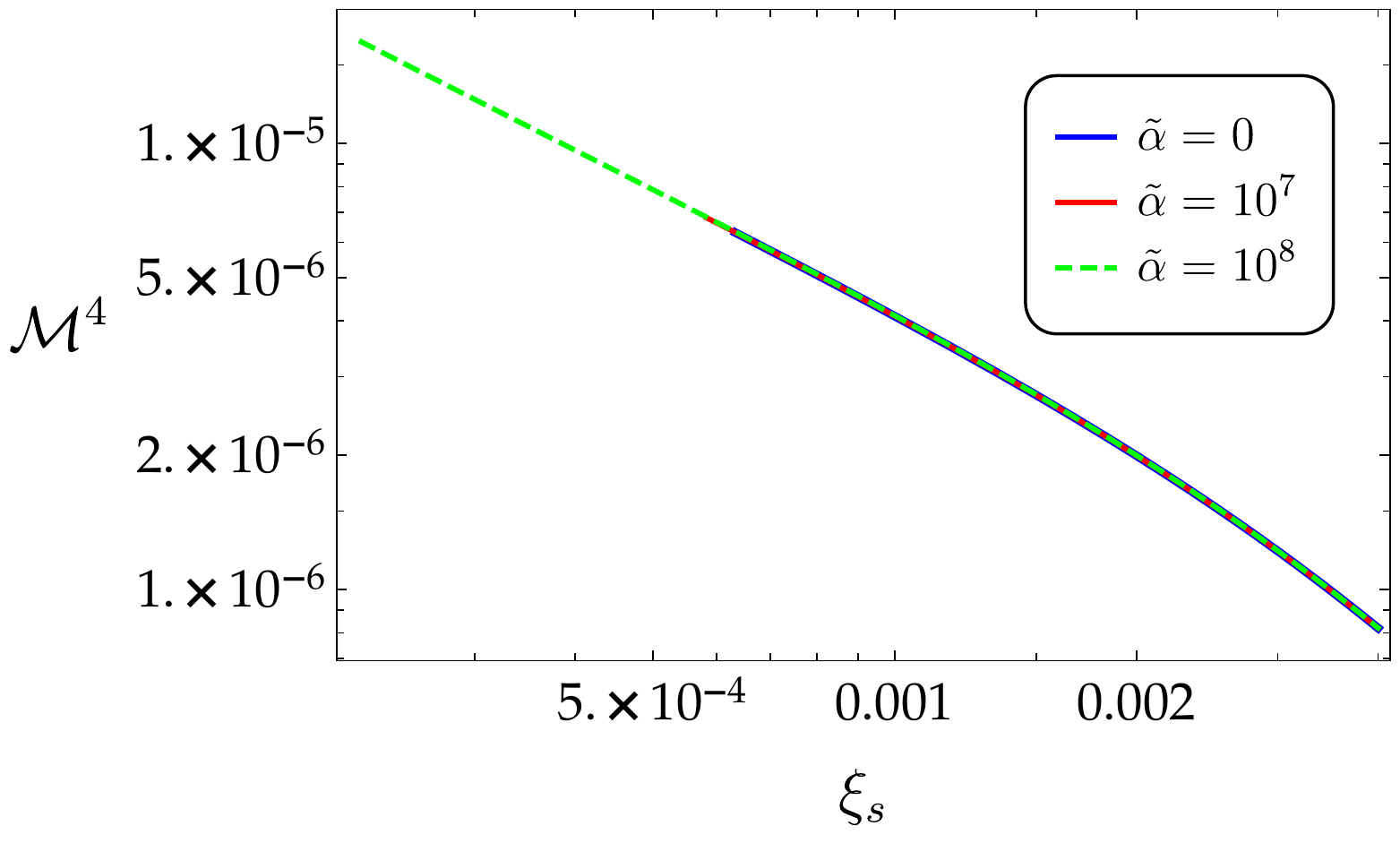}
\caption{\sf The parameter $\mathcal{M}^4$ as a function of $\xi_{\rm s}$, for the viable range of values for $\xi_{\rm s}$ as given in Table~\ref{table:viable_xi_s_small_field} for small field inflation. For $\ta \gtrsim 10^{8.267} \simeq 1.85 \times 10^{8}$, the $\xi_{\rm s} \rightarrow 0$ limit is located in the observationally viable $95\%$ CL region of the $r$-$n_s$ plot and thus there is no lower cutoff for the value of $\xi_{\rm s}$. Consequently in this case there is no upper cutoff for $\mathcal{M}^4$.}
\label{fig:xi_s_vs_M_4}
\end{figure}

A further inspection of Fig.~\ref{fig:r_ns} reveals that for values of $\ta \lesssim 10^{8.267} \simeq 1.85 \times 10^{8}$, LFI is not viable since its predictions lay outside the $95\%$ CL region for the measured values for $r$ and $n_s$. On the other hand, SFI complies with observations for a finite range of values of $\xi_{\rm s}$ with the smallest (largest) viable value of $\xi_{\rm s}$ yielding the largest (smallest) predicted value for $r$ for a given $\ta$, see also Table~\ref{table:viable_xi_s_small_field}. This range is $2\times 10^{-4} \lesssim \xi_s \lesssim 4\times 10^{-3}$ and consequently, via~\eqref{eq:xi_svev} the VEV of the inflaton is restricted to $15\, M_{\rm P} \lesssim v_s \lesssim 70\, M_{\rm P}$. Furthermore, the finite range of allowed values for $\xi_{\rm s}$ implies a corresponding finite range of viable values for the parameter $\mathcal{M}$ as can be seen in Fig.~\ref{fig:xi_s_vs_M_4}. 

\begin{table}
\begin{center}
\begin{tabular}{| c | c | c | c | c |c |}
\hline \hline
\multicolumn{6}{c} { Small field inflation }    \\ 
\hline \hline
%\multicolumn{6}{c} { Minimum viable $\xi_{\rm s}$ values }    \\      
%\hline \hline  
$\ta$ & $ \xi_{\rm s}^{\rm (min)} $ & $ \mathcal{M} $ & $ r $ & $ n_s $ & $ N_\star $  \\ \hline
 & & & & \\[-1em]
$ 0 $ & $ 0.0006267 $ & $ 0.0502432 $ & $ 0.0729636 $ & $ 0.968159 $ & $ 60.3 $ \\ 
\hline
$ 10^{7} $ & $ 0.0005830 $ & $ 0.0510926 $ & $ 0.0730490 $ & $ 0.968233 $ & $ 60.3 $ \\
\hline
$ 10^{8} $ & $ 0.0002017 $ & $ 0.0651665 $ & $ 0.0732724 $ & $ 0.968439 $ & $ 60.3 $ \\
\hline \hline 
%\multicolumn{6}{c} { Maximum viable $\xi_{\rm s}$ values }   \\    
%  \hline \hline   
 $\ta$ & $ \xi_{\rm s}^{\rm (max)} $ & $ \mathcal{M} $ & $ r $ & $ n_s $ & $ N_\star $  \\ \hline 
  & & & & \\[-1em]
$ 0 $ & $ 0.0041417 $ & $ 0.0297085 $ & $ 0.0161109 $ & $ 0.957741 $ & $ 59.6 $ \\ 
\hline
$ 10^{7} $ & $ 0.0041389 $ & $ 0.0297168 $ & $ 0.0160355 $ & $ 0.957747 $ & $ 59.6 $ \\
\hline 
$ 10^{8} $ & $ 0.0041367 $ & $ 0.0297308 $ & $ 0.0152745 $ & $ 0.957739 $ & $ 59.6 $ \\
\hline \hline 
\end{tabular}
\caption{\sf For $\ta \lesssim 10^{8.267} \simeq 1.85 \times 10^{8}$, only small field inflation yields viable values for $r$ and $n_s$ (see Fig.~\ref{fig:r_ns}). Here, we give the minimum and maximum values of $\xi_{\rm s}$ for which we obtain viable predictions for various $\ta$. We also give the values of $\mathcal{M}$, $r$, $n_s$ and $N_\star$ for these marginal values of $\xi_{\rm s}$.} \label{table:viable_xi_s_small_field}
\end{center}
\end{table}

For values of $\ta \gtrsim 10^{8.267} \simeq 1.85 \times 10^{8}$ the $\xi_{\rm s} \rightarrow 0$ limit is located within the observationally viable $95\%$ CL region of the $r$-$n_s$ plot (see Fig.~\ref{fig:r_ns}) and so SFI and LFI exhibit only an upper cutoff, $\xi_s \lesssim 4\times 10^{-3}, $ for the viable values of $\xi_{\rm s} $ as it is shown in Table~\ref{table:Large_small_xi_s_max}. This in turn implies a lower cutoff, $15\, M_{\rm P} \lesssim v_s$, for the VEV of the inflaton. 

To summarize, in all the cases $v_s$ must be super-Planckian which imposes that $v_s \simeq v_\phi$ as $v_h \sim \mathcal{O}(10^{-16}) \, M_{\rm P}$. It is then evident that the mixing angle as defined in Eq.~\eqref{eq:FD_angle_def} will satisfy $\omega \simeq 0$ and thus the flat direction, for values of the parameters that lay in the viable regions of the parametric space, will be nearly identified with the direction of the field $\phi$ in field space, see Fig.~\ref{fig:Potentials}.

\begin{table}
\begin{center}
\begin{tabular}{| c | c | c | c | c |c |}
\hline \hline
\multicolumn{6}{c} {Small field inflation }   \\    
  \hline \hline   
 $\ta$ & $ \xi_{\rm s}^{\rm (max)} $ & $ \mathcal{M} $ & $ r $ & $ n_s $ & $ N_\star $  \\ \hline 
  & & & & \\[-1em]
$ 10^{9} $ & $ 0.0040967 $ & $ 0.0299033 $ & $ 0.0103843 $ & $ 0.957734 $ & $ 59.4 $ \\ \hline  
$ 10^{10} $ & $ 0.0039033 $ & $ 0.0306853 $ & $ 0.0024263 $ & $ 0.957835 $ & $ 58.8 $ \\ \hline  
$ 10^{11} $ & $ 0.0036767 $ & $ 0.0316817 $ & $ 0.0002763 $ & $ 0.957919 $ & $ 58.0 $ \\ \hline  
$ 10^{12} $ & $ 0.0035200 $ & $ 0.0324432 $ & $ 0.0000280 $ & $ 0.957921 $ & $ 57.3 $ \\
\hline \hline
\multicolumn{6}{c} {Large field inflation }    \\      
\hline \hline  
$\ta$ & $ \xi_{\rm s}^{\rm (max)} $ & $ \mathcal{M} $ & $ r $ & $ n_s $ & $ N_\star $  \\ \hline
 & & & & \\[-1em]
$ 10^{9} $ & $ 0.0028733 $ & $ 0.0245332 $ & $ 0.0248280 $ & $ 0.958142 $ & $ 60.0 $ \\ \hline 
$ 10^{10} $ & $ 0.0025667 $ & $ 0.0259176 $  & $ 0.0027631 $ & $ 0.957819 $ & $ 59.0 $\\ \hline 
$ 10^{11} $ & $ 0.0020250 $ & $ 0.0286194 $ & $ 0.0002796 $ & $ 0.957922 $ & $ 58.1 $ \\ \hline  
$ 10^{12} $ & $ 0.0017108 $ & $ 0.0306805 $ & $ 0.0000280 $ & $ 0.957918 $ & $ 57.4 $ \\ 
\hline \hline 
\end{tabular}
\caption{\sf For various $\ta \gtrsim 10^{8.267} \simeq 1.85 \times 10^{8}$, and for both small and large field inflation, we give the corresponding maximum values of $\xi_{\rm s}$ that yield predictions that comply with the observational bounds. We also give the values of $\mathcal{M}$, $r$, $n_s$ and $N_\star$ for these marginal values of $\xi_{\rm s}$.} \label{table:Large_small_xi_s_max}
\end{center}
\end{table}

%%%%%%%%%%%%%%%%%%%%%%%%%%%%%%%%%%%%%%%%%%%%%%
\section{Conclusions}
\label{sec:Conclusions}

In this paper we have studied  a model of scale-invariant quadratic gravity in the context of the Palatini formulation. The Planck scale is  dynamically generated via the Coleman-Weinberg formalism through the VEVs of the scalar field $\phi$ and the Higgs field $h$. These scalar fields were nonminimally coupled to gravity through terms of the form $\xi_i \Phi_i^2 R$, where $\Phi_i =\phi, h$. The extra scalar field $\phi$ originated from an $U(1)_X$ extension of the SM containing an extra gauge boson $X_\m$ and three right-handed neutrinos $N^i_R$. The Higgs mass was generated through the portal coupling $\lambda_{h \phi} h^2 \phi^2$. This is exactly the significance of the addition of the extra scalar field $\phi$. Without it, the necessity of the existence of a Higgs mass term with a dimensionful coupling would have broken the scale invariance of our model. A possible extra $Z_2$ symmetry facilitates the stability of the potential dark matter candidates in the context of our model. As discussed, these can be either the new fermions of the model e.g. the right-handed neutrinos or the extra Dirac fermion $\zeta$, or the extra $U(1)_X$ gauge boson.

We have  employed  the Gildener-Weinberg approach, the  generalization of the Coleman-Weinberg mechanism to the multiple fields case, in order to identify the flat direction of the tree-level potential. Along the flat direction, the theory effectively becomes single field and by computing the quantum corrections we obtain the one-loop effective potential, which is stabilized due to the extra $U(1)_X$ gauge boson. In the effective single-field description, two parameters are important for our analysis namely the effective nonminimal coupling $\xi_s$, which is constructed out of the nonminimal couplings of $\phi$ and $h$ and their mixing angle $\omega$, and the effective higher-curvature coupling $\ta$ which corresponds to a combination of the coupling constants of the quadratic curvature corrections in the action. These quadratic in curvature terms are the usual scale invariant terms $R^2$ and $R_{(\m\n)}R^{(\m\n)}$. The fact that their effect on the inflationary observables can be described collectively by the common coupling $\ta$ reveals that their contribution to the final EF potential is the same. 
On the other hand, the higher order kinetic terms generated in the EF are not of the same form, as the $R^2$ term gives us only a second order kinetic term, while the $R_{(\m\n)}R^{(\m\n)}$ term gives higher than the second order terms. The study of such kinetic terms was out of the scope of this paper as they are negligible at least during slow roll. 

In order to transform the action into the EF and compare the predictions of the model with observations, the use of both conformal and disformal transformations is required. The one-loop corrections are taken in the ``intermediate frame", that is after having performed the conformal transformation that decouples the scalar fields from the Einstein-Hilbert term, but before the disformal transformation that removes the quadratic curvature terms from the gravity sector. It is in this intermediate frame that we may have a one-loop effective potential with a minimum at zero without invoking a cosmological constant term that would render our model \enquote{quasi scale invariant}.
Upon recasting the action to the EF, we end up with a modified effective potential $\ubar(s_c)$ in terms of a canonical scalar field $s_c$ that plays the role of the inflaton. The shape of the potential $\ubar(s_c)$ exhibits plateaus on both sides of the minimum and thus both small field inflation (SFI) and large field inflation (LFI) can be accommodated in our model. The additional higher-order kinetic terms that arise in the EF are negligible in the slow roll approximation, and so we have retained only linear order terms in our analysis. Applying the cosmological data on inflation we were able to constrain the size of the VEV of these scalar fields, and consequently the masses of the extra gauge boson and the right-handed neutrinos.

In order to constrain the parametric space we have considered the latest bounds on cosmological observables as set by the Planck collaboration and we have found that our model complies with observations for a wide range of parameters. More precisely, for values of the parameter $\ta$ in the range  $\ta \gtrsim 1.85 \times 10^{8}$, both SFI and LFI support viable inflation when $\xi_s \lesssim \mathcal{O}(10^{-3})$. In the large $\ta$ limit, the inflationary potential becomes symmetric about its minimum and consequently the predictions for the observables of SFI and LFI are identical.

When $\ta \lesssim 1.85 \times 10^{8}$, and independently of the value of $\xi_s$, LFI is nonviable since the predicted values for the tensor-to-scalar ratio and the tilt of the scalar power spectrum lay outside the $95\%$ CL region. On the other hand, SFI exhibits regions in the parametric space that are viable for any $\ta$ with $\xi_s$ interpolating between a maximum and a minimum value. Eventually, the largest viable value for $\xi_s$ in our model is obtained within the context of SFI and is approximately $\xi_s \simeq 4 \times 10^{-3}$ which translates to a minimum value for the VEV of $s_c$ in the vicinity of $15\, M_{\rm P}$. It will be interesting to investigate whether the rest of the shortcomings of the SM, such as the strong CP problem, can be addressed in a similar setting where successful inflation can be realized and the dark matter and baryon asymmetry problems can be solved in a common framework.

%-------------------------------------------------------------------------------
\acknowledgments
%-------------------------------------------------------------------------------
 We would like to thank A. Racioppi for useful discussions. The research of IDG is co-financed by Greece and the European Union (European Social Fund- ESF) through the Operational Programme \textquote{Human Resources Development, Education and Lifelong Learning} in the context of the project \textquote{Strengthening Human Resources Research Potential via Doctorate Research - 2nd Cycle} (MIS-5000432), implemented by the State Scholarships Foundation (IKY). AK was supported by the Estonian Research Council grants MOBJD381 and MOBTT5 and by the EU through the European Regional Development Fund CoE program TK133 ``The Dark Side of the Universe." TDP acknowledges the support of the grant 19-03950S of Czech Science Foundation (GA\v{C}R). Finally,  the research work of VCS was supported by the Hellenic Foundation for Research and Innovation (H.F.R.I.) under the ``First Call for H.F.R.I. Research Projects to support Faculty members and Researchers and the procurement of high-cost research equipment grant'' (Project Number: 824). 

%\vspace{3cm}

%\newpage
\appendix
%%%%%%%%%%%%%%%%%%%%%%%%%%%%%%%%%%%%%%%%%%%%%%
\section{Details on the variations }
\label{sec:appendixA}
 Substituting~\eqref{eq:curvfun} in~\eqref{eq:qmn} gives us that the auxiliary field $\Sigma_{\m\n}$ in terms of $q_{\m\n}$ and $g_{\m\n}$ reads
\be
\Sigma_{\m\n}=\frac{1}{2\b}\frac{\sqrt{-q}}{\sqrt{-g}} q^{\k\l} g_{\k\m} g_{\l\n}-\frac{1}{2\b+8\a}\left(1+\frac{\a}{\b} \frac{\sqrt{-q}}{\sqrt{-g}} q^{\k\l}g_{\k\l}\right)g_{\m\n}\,,
\label{eq:sigmamn}
\ee
and its trace is
\be
\Sigma=g^{\m\n} \Sigma_{\m\n}=\frac{-4}{2\b+8\a} + \frac{1}{2\b+8\a} \frac{\sqrt{-q}}{\sqrt{-g}}q^{\k\l}g_{\k\l}\,.
\label{eq:sigma}
\ee
The part of the Lagrangian density that has to be varied with respect to $g_{\m\n}$ is
\ba
\sqrt{-g}\mathcal{L}_g &=& - \sqrt{-g} \left( \frac{1}{2} \frac{\partial C}{\partial \Sigma_{\m\n}} \Sigma_{\m\n} -\frac{1}{2} C -\mathcal{L}_m \right) \nonumber
\\ &=& - \sqrt{-g} \left( \frac{\a}{2}\Sigma^2 +\frac{\b}{2} \Sigma_{\m\n} \Sigma^{\m\n} + \frac{1}{2} g^{\m\n} \partial_\m s_c \partial_\n s_c  + U_{\rm eff}(s_c) \right)\,.
\label{eq:lagG}
\ea
Varying~\eqref{eq:lagG} we obtain that
\ba
\d\left(\sqrt{-g} \mathcal{L}_g\right) &=& - \sqrt{-g}\bigg[ \a \Sigma \d\Sigma +\b g^{\m\g}\Sigma_{\m\n}\d\left(g^{\r\n}\Sigma_{\g\r}\right) + \frac{1}{2} \partial_\m s_c \partial_\n s_c \d  g^{\m\n} \bigg] \nonumber
\\ && -\frac{1}{2}\sqrt{-g}g_{\m\n}\d g^{\m\n}\bigg[-\frac{\a}{2}\Sigma^2 -\frac{\b}{2}\Sigma_{\m\n}\Sigma^{\m\n}-U_{\rm eff}(s_c) - \frac{1}{2}g^{\k\l} \partial_\k s_c \partial_\l s_c  \bigg]\,.
\label{eq:dlagGvar}
\ea
Substituting~\eqref{eq:sigmamn} and~\eqref{eq:sigma} in~\eqref{eq:dlagGvar} and after manipulations we have that
\ba
\frac{1}{\sqrt{-g}}\frac{\d S}{\d g^{\m\n}} = && -\frac{1}{4(\b+4\a)}\frac{\sqrt{-q}}{\sqrt{-g}}q^{\s\l}g_{\s\m}g_{\l\n} \nonumber
\\&& +\frac{1}{4\b}\frac{q}{g}\left(q^{\s\l}q^{\r\d}g_{\l\d}g_{\r\n}g_{\s\m} -\frac{\a}{\b+4\a}q^{\d\r}g_{\d\r}q^{\s\l}g_{\s\m}g_{\l\n} \right) \nonumber
\\&& +\frac{1}{2}g_{\m\n} \left[\frac{1}{\b+4\a}\left( \frac{1}{2}+\frac{\a}{8\b}\frac{q}{g}q^{\l\s}g_{\l\s}q^{\r\d}g_{\r\d}\right) -\frac{q}{g}\frac{1}{8\b}q^{\l\s}q^{\d\r}g_{\l\d}g_{\s\r}\right] \nonumber
\\ && + \frac{1}{2}g_{\m\n} \left(\frac{1}{2}g^{\l\s} \partial_\l s_c \partial_\s s_c  +  U_{\rm eff}(s_c) \right) - \frac{1}{2}\partial_\m s_c \partial_\n s_c =0\,.  
\label{eq:dactionAP}
\ea

%%%%%%%%%%%%%%%%%%%%%%%%%%%%%%%%%%%%%%%%%%%%%%
\section{The functions $R_i$, $a_i$,  and $b_i$ }
\label{sec:appendixB}
The functions $R_i$ which have been displayed in Eqs.~\eqref{eq:eq1}-\eqref{eq:eq2} are listed below
\ba
 R_1 &=& B\left(2A-2B X_q \right) \,, \nonumber
\\   R_2 &=& 4A -2B X_q \,, \nonumber
\\   R_3 &=& 4A^2 -4AB  X_q  +4B^2  X_q^2 \,,  \nonumber
\\  R_4 &=& A(R_1+AB) -2 BR_1 X_q  \,, \nonumber
\\   R_5 &=& A^3\left(A-2B  X_q \right)\,.
\label{eq:Ri}
\ea
The coefficients $a_i$,  and $b_i$ which have been displayed in Eq.~\eqref{eq:solslowroll} are\footnote{These coefficients have been also found in~\cite{Annala2020}.}
\ba
a_0 &=& \frac{1}{1+\ta U_{\rm eff}}\,,  \nonumber
\\ b_0 &=& \frac{ (\tb-\ta)}{ (1+\ta U_{\rm eff}) (1+\tb U_{\rm eff})}\,, \nonumber
\\ a_{1} &=& \frac{\tb}{2  (1+\tb U_{\rm eff})}\,,  \nonumber
\\ b_{1} &=& \frac{(\tb-\ta) \left(3 \tb-2 \ta + (2 \tb-\ta) (\ta+\tb)U_{\rm eff} +\ta \tb^2 U_{\rm eff}^2\right)}{ (1+\ta U_{\rm eff}) (1+\tb U_{\rm eff})^3}\,, 
\label{eq:ai}
\ea
where we have defined $\ta=2\b+8\a$, $\tb=4\b+8\a$ and $U_{\rm eff}=U_{\rm eff}(s_c)$.  As it seems, for $\ta=\tb$, the coefficients $b_0$ and $b_1$, like  the rest higher order $b$ coefficients of the same series~\eqref{eq:solslowroll}, are equal to zero. This is expected, as the equality of the tilted factors is translated to an elimination of the $R_{\m\n}R^{\m\n}$ term and so the disformal transformation is reduced again to the usual conformal.

%\vspace{1 cm}

\bibliography{References}{}

\end{document}